\documentclass{amsart}%[10pts]
\usepackage{NCmacrosfp}
\begin{document}
\title[$\mathcal{M}$-Decomposability]
{$\mathcal{M}$-Decomposability, Elliptical Unimodal Densities, and 
Applications to Clustering and Kernel Density Estimation}
\author[Nicholas Chia]{Nicholas Chia}
\address{The Institute of Statistical Mathematics,
10-3 Midori-cho, Tachikawa, Tokyo 190-8562, Japan.}
\email{sha@ism.ac.jp (Nicholas Chia)}
\author[Junji Nakano]{Junji Nakano}
\address{The Institute of Statistical Mathematics,
10-3 Midori-cho, Tachikawa, Tokyo 190-8562, Japan.}
%
%\begin{document}
%
\begin{abstract}
\cite{ChiaNakano1d} introduced the
concept of $\mathcal{M}$-decom-posability of probability densities in
one-dimension. 
In this paper, we generalize $\mathcal{M}$-decomposability to
any dimension. We prove that all elliptical unimodal densities 
are $\mathcal{M}$-undecomposable. We also derive an inequality to 
show that it is better to represent an $\mathcal{M}$-decomposable density via
a mixture of unimodal densities.
Finally, we demonstrate the application of $\mathcal{M}$-decomposability to
clustering and kernel density estimation, using real and simulated data. Our
results show that $\mathcal{M}$-decomposability can be used as a non-parametric
criterion to locate modes in probability densities.

%{\bf keywords:}
%{\it covariance matrices, cluster analysis, 
%elliptical unimodal densities, inequality, kernel density estimation, 
%Kullback-Leibler divergence, $\mathcal{M}$-decomposability, mode location, 
%non-parametric criterion}
\end{abstract}

\maketitle
\section{Introduction}
\label{sec:introduction}
In a recent paper, \cite{ChiaNakano1d} conceptualized 
$\mathcal{M}$-decomposability and developed the theory in one-dimension. 
The main results are summarized in the following paragraph.

$\mathcal{M}$-decomposability is defined as follows. Let $f$ be a probability 
density defined in one-dimension. There exist countless ways to express $f$ as 
a weighted mixture of two probability densities, in the form of
\[
f(x) = \alpha \, g(x) + (1-\alpha) \, h(x) \quad \text{  where } 0 < \alpha < 1 \,.
\]
If it is possible to find any combination of $\{\alpha,g,h\}$, which satisfies
\[
\sigma_{f} >
\sigma_{g} + \sigma_{h} \quad \text{where } \sigma_{f} 
\text{ denotes the standard deviation of } f,
\] 
then the original density $f$ is said to be {\it $\mathcal{M}$-decomposable}. 
Otherwise, $f$ is {\it $\mathcal{M}$-undecomposable}. Intuitively, multimodal 
densities with peaks separated far apart are likely to be 
$\mathcal{M}$-decomposable. Conversely, unimodal densities are probably 
$\mathcal{M}$-undecomposable. The authors proved that all one-dimensional 
{\it symmetric unimodal densities} with finite second moments are 
$\mathcal{M}$-undecomposable. In other words, if $f$ is symmetric unimodal and 
has finite second moments, then for any weighted mixture density components 
$\{g,h\}$ of $f$, one must have
\begin{equation}
\label{eq:inequality1}
\sigma_{f} \leq \sigma_{g} + \sigma_{h} \,.
\end{equation}
Eq~(\ref{eq:inequality1}) applies to a wide range of densities that include 
Gaussian, Laplace, logistic and many others. The authors also showed the 
possibility of using $\mathcal{M}$-decomposability to perform cluster analysis 
and mode finding in one-dimension. Incidentally, the ``$\mathcal{M}$" in 
$\mathcal{M}$-decomposability may either mean ``multimodal" or ``mixture".

In this paper, we further contribute to $\mathcal{M}$-decomposability, both in 
the theoretical and applicational aspects. On the theoretical front, we 
generalize the concept of $\mathcal{M}$-decomposability to any $d$-dimensional 
space. First of all, we derive a theorem (Theorem~\ref{th:maind}) 
that is the $d$-dimensional equivalent of Eq~(\ref{eq:inequality1}).
We prove that all {\em elliptical unimodal densities} with finite second
moments are $\mathcal{M}$-undecomposable. These densities include multivariate 
Gaussian, Laplace, logistic and many others. Following that, we derive another 
theorem, (Theorem~\ref{th:KLD}), which determines if a given density is better 
approximated via a mixture of Gaussian densities, instead of one single 
Gaussian density.

One example of application of $\mathcal{M}$-undecomposability is cluster
analysis. For decades, cluster analysis has been a popular research subject, 
both from the theoretical and algorithmic aspects. Cluster analysis is likely to
remain a widely researched topic, given the many different approaches that
caters to varying applications. The survey paper by \cite{Berkhin} provides an 
up-to-date status of available clustering techniques and methodologies. 
There are two main classes of cluster analysis methodologies: 
{\it parametric} and {\it non-parametric}. For parametric cluster analysis, 
one needs prior knowledge or assumptions on the analytical structure of the 
underlying clusters. The whole dataset is modeled as a mixture of $k$ 
parametrized densities, and the problem reduces to parameter estimation. 
In \cite{McLachlanPeel}, parametric cluster analysis via the 
{\it Expectation-Maximization} (EM) algorithm is described in detail. 
Other parametric methods include the Bayesian particle filter approach 
detailed in \cite{Fearnhead}, and the reversible jump 
{\it Markov chain Monte Carlo} (MCMC) approach by \cite{RichardsonGreen}. 
For parametric cluster analysis, the most popular approach is to model the
clusters as Gaussian densities. 

As for non-parametric cluster analysis, a popular tool is the $k$-means 
algorithm. The $k$-means algorithm is optimal for locating similar-sized 
spherical clusters within a dataset, provided the number of clusters are known
beforehand. %As such, the $k$-means algorithm is not robust to scaling. 
With elliptical clusters, or clusters of varying sizes, the
$k$-means approach yields results that are meaningless.
The $k$-means algorithm assigns samples to clusters based on {\it distance}
(Euclidean or its variations) to the centres of the clusters. Other 
{\it distance-based} non-parametric clustering algorithms include the 
nearest-neighbour clustering. Distance-based clustering algorithms generally
share the same drawbacks such as sensitivity to scaling, elliptical clusters and
clusters of varying sizes. If the number of clusters are not known 
beforehand, neither the $k$-means algorithm nor the nearest-neighbour algorithm
 estimate the number of clusters automatically. For the $k$-means algorithm, 
 the unknown number of clusters has to be re-evaluated via 
 {\it Akaike's information criterion} (AIC), proposed by \cite{Akaike}, 
or other suitable model selection criterion. 

Our approach to cluster analysis via $\mathcal{M}$-decomposability is 
non-parametric and are based on {\it volume} instead of distance.
Being non-parametric, prior knowledge on the analytical structure of the 
underlying clusters is unnecessary. The only assumption required is that the 
clusters are approximately elliptical and unimodal. As a result, the 
limitation of clustering via $\mathcal{M}$-undecomposability is that it will 
probably not perform ideally for irregularly shaped clusters that deviate from 
elliptical unimodal densities. However, if the clusters are approximate 
elliptical and unimodal, then our clustering methodology works well, and allows 
for the unknown number of clusters to be recovered automatically. Furthermore,
as clustering via $\mathcal{M}$-decomposability is based on volume instead of
distance, cluster allocation is invariant to scaling.

For existing alternative methodologies to clustering, there has been recent
development on Rousseeuw's minimum volume ellipsoids (MVE) in
\cite{RousseeuwLeroy} and \cite{RousseeuwZomeren}. The MVE approach 
is originally developed as a robust method to estimate mean vectors and 
covariance matrices of multivariate data in the presence of outliers.
MVE is computationally intensive and the optimal solution is often difficult
to achieve, prompting many research papers on the algorithmic aspects of the
problem. Some authors, for example, \cite{ShiodaTuncel}, 
outlined a heuristic for clustering via MVE by minimizing the sum of volume of clusters.
Our methodology of clustering via $\mathcal{M}$-decomposability has some 
similarities with clustering via the MVE 
approach, in that both measure ``volume" in a certain sense.
Central to the $\mathcal{M}$-decomposability concept is the ``pseudo-volume",
which we define as the square-root of the determinant of the covariance matrix.
Compared to MVE, the pseudo-volume is computationally cheap and straightforward. 
On top of that, we also provide theoretical justifications in 
Theorem~\ref{th:KLD} for minimizing the sum of pseudo-volumes of clusters.

Another possible area of application of $\mathcal{M}$-undecomposability 
is density estimation. In density estimation, data 
generated from some unknown densities are given, and 
the task is to estimate and recover the unknown density. 
One popular non-parametric approach to density
estimation is kernel density estimation, 
treated in \cite{Silverman}, \cite{Scott}, \cite{Hardle},
as well as \cite{WandJones}. The difficulty in kernel density estimation is the
derivation of the optimal kernel bandwidth: If the kernel bandwidth is
underestimated, the kernel density becomes unduly spiky; if the kernel
bandwidth is overestimated, the kernel density becomes oversmoothed. 
For multimodal densities, it is not possible to find a single kernel bandwidth
that provides a satisfactory density estimation everywhere. 
Using $\mathcal{M}$-decomposability, we demonstrate that there is a simple and
logical way to circumvent the above problem by representing the underlying 
density as a mixture of unimodal densities where necessary.

This paper develops both the theoretical and applicational aspects of
$\mathcal{M}$-decom-posability, and therefore should be of interest to
theoretical statisticians and practitioners alike.
Section~\ref{sec:MDd} is devoted to the theoretical development of
$\mathcal{M}$-decomposability in $d$-dimensional space.
For readers who are only interested in applications, it is possible to note only 
the results of Theorems~\ref{th:maind} and \ref{th:KLD}, skipping the rest of
Section~\ref{sec:MDd} without disrupting the flow of the paper.

\section{$\mathcal{M}$-Decomposability in $d$-Dimensional Space}
\label{sec:MDd}

\subsection{Extensions from One-Dimension}
In \cite{ChiaNakano1d}, $\mathcal{M}$-decom-posability involves only
the standard deviations of probability densities. This is because
 in one-dimension, the standard deviation is a natural
measure of scatter of a given density. The standard 
deviation of any density in one-dimension has the same order as the 
distance or ``length" computed from the mean. 
When considering higher dimensions, a possible corresponding measure of scatter
of a given density is
the square-root of the determinant of the covariance matrix of the density.
The square-root of the determinant of the covariance matrix in $d$-dimensional
space has the same order as $d$-dimensional ``hypervolume".
Henceforth, we shall call the above measure the {\it pseudo-volume} of a
density. We denote the covariance matrix of a
density $f$ by $\Sigma_{f}$, and therefore the pseudo-volume of $f$ is given
by $|\Sigma_{f}|^{\frac{1}{2}}$.
In one-dimension, pseudo-volume reduces to the standard deviation.

In \cite{ChiaNakano1d}, 
the authors limited the number of mixture components to two in their 
development of $\mathcal{M}$-decomposability.
In this paper, we show that it is possible to relax the above limitation, and 
generalize the number of mixture components to $m$ where
$m \geq 2$. %density subcomponents.
Let $f$ be a probability density function defined on $\mathcal{R}^{d}$, the
$d$-dimensional real space. 
One can always express $f$ as a weighted mixture
of $m$ densities as follows: %$\{g_{1}, \ldots, g_{m}\}$ as follows:
\begin{equation}
\label{eq:dp}
f({\bf x}) = \alpha_{1} \, g_{1}({\bf x}) + \cdots +
\alpha_{m} \, g_{m}({\bf x}) \,,
\end{equation}
where $0 < \alpha_{i} < 1$ and $\Sigma \, \alpha_{i} = 1$.
Henceforth, we call any set of densities $\{g_{1}, \ldots ,g_{m}\}$ 
which satisfies Eq~(\ref{eq:dp}) a set of {\em mixture components} of $f$.

We extend the definition of 
$\mathcal{M}$-decomposability to $d$-dimensional space as follows.
\begin{definition}[$\mathcal{M}$-Decomposability]
\label{def:MDd}
For a given probability density function $f$, if there exists a set of mixture
components $\{g_{1}, \ldots, g_{m}\}$ such that
\[
|\Sigma_{f}|^{\frac{1}{2}} > |\Sigma_{g_{1}}|^{\frac{1}{2}} + \ldots 
+ |\Sigma_{g_{m}}|^{\frac{1}{2}} \, ,
\]
then $f$ is defined to be {\it $\mathcal{M}$-decomposable}. 
Otherwise, $f$ is
{\it $\mathcal{M}$-undecomposable}. 
If for any set of mixture
components $\{g_{1}, \ldots, g_{m}\}$,
\[
|\Sigma_{f}|^{\frac{1}{2}} < |\Sigma_{g_{1}}|^{\frac{1}{2}} + \ldots 
+ |\Sigma_{g_{m}}|^{\frac{1}{2}} \, ,
\] 
then $f$ is {\it strictly $\mathcal{M}$-undecomposable}.
\end{definition}
%
%\begin{remark}
Our new definition of $\mathcal{M}$-decomposability reduces to that presented 
in \cite{ChiaNakano1d} when $m=2$ and $d=1$. For $d \geq 2$,
the definition of $\mathcal{M}$-decomposability can be described compactly using
pseudo-volumes. %(Refer to Definition~\ref{def:pv}.)
%\end{remark}
 % \label{def:MDd}

\subsection{Elliptical Uniform Densities}
The uniform density is trivially defined in one-dimension, but in higher
dimensions, it may assume many different possible shapes. 
For example, one may think of the uniform hypercube or the uniform hypersphere. 
However, the subject of interest in our paper is
the {\it elliptical uniform density}, which forms the fundamental building 
block of elliptical unimodal densities. 

Ellipticity, uniformity and unimodality are three different qualities. The
definitions of the first two are given immediately below, and the third will be 
given in Section~\ref{sec:EUD}.

\begin{definition}[Elliptical and Spherical Densities]
\label{def:spherical}
We say that $f$ is {\em elliptical} if there exist a vector $\mu \in
\mathcal{R}^{d}$, a positive semidefinite symmetric matrix 
$\Sigma \in \mathcal{R}^{d \times d}$ and a
positive function $p$ on $\mathcal{R}^{+} \cup \{0\}$ such that
\[
f({\bf x}) = p\{({\bf x} - \mu)^{T} \, \Sigma^{-1} \, ({\bf x} - \mu)\} \,.
\]
Furthermore, if $\Sigma = k \,
\mathbf{I}_{d}$, where $k>0$ and $\mathbf{I}_{d}$ denotes the $d$-dimensional 
identity matrix, then $f$ becomes 
\[
f({\bf x}) = p_{1}\{({\bf x} - \mu)^{T} \, ({\bf x} - \mu)\} 
= p_{2}(|{\bf x} - \mu|) \,,
\]
and we say that $f$ is {\em spherical}.
\end{definition}
\begin{remark}
The mean and covariance matrix of the above-defined elliptical density $f$ are
as follows:
\[
\mu_{f} = \mu, \quad \Sigma_{f} = c \, \Sigma \quad \text{where } c > 0 \,.
\] %where $c > 0$.
\end{remark}

\begin{definition}[Uniform Densities]
\label{def:Uniform}
We say that $f$ is {\em elliptical uniform} if there exist a vector $\mu \in
\mathcal{R}^{d}$, a positive semidefinite symmetric matrix 
$\Sigma \in \mathcal{R}^{d \times d}$, and a positive real number $r$ such that
\[
%\begin{equation}
%\label{eq:ellipticalUniform}
f({\bf x}) \propto 
\mathbb{I}_{({\bf x}-\mu)^{T} \, \Sigma^{-1} \, ({\bf x}-\mu)<r^{2}} \, ,
%\end{equation}
\]
where $\mathbb{I}$ denotes the indicator function.
Furthermore, if $\Sigma = k \,
\mathbf{I}_{d}$, where $k>0$ and $\mathbf{I}_{d}$ denotes the $d$-dimensional 
identity matrix, then $f$ becomes 
\[
%\begin{equation}
%\label{eq:sphericalUniform}
f({\bf x}) \propto 
\mathbb{I}_{({\bf x}-\mu)^{T} \, ({\bf x}-\mu) < {r^{\prime}}^{2}} \, 
= \mathbb{I}_{|{\bf x}-\mu| < r^{\prime}} \, ,
%\end{equation}
\]
and we say that $f$ is {\em spherical uniform}.
\end{definition}

\begin{theorem}[Inequality on Elliptical Uniform Densities]
\label{th:ud}
All elliptical uniform densities defined on $\mathcal{R}^{d}$ are 
$\mathcal{M}$-undecomposable
in $d=1$ and strictly $\mathcal{M}$-undecomposable for $d \geq 2$.
\end{theorem}
 % \label{th:ud}

The proof of Theorem~\ref{th:ud} proceeds the following lemma.

\begin{lemma}[Density with Minimum Pseudo-volume]
\label{lm:MinDetCovd}
Let $f$ be a probability density function defined on 
${\bf x} \in {\mathcal R}^{d}$ such that
$f({\bf x}) \leq M_{f}$ for all ${\bf x}$. Then 
%\in {\mathcal R}^{d}$, or 
%$\max (f) \leq M_{f} < \infty$. Then
\[
|\Sigma_{f}|^{\frac{1}{2}} \geq
\frac{\Gamma(\frac{d}{2}+1)}
{M_{f} \, \{\pi \, (d+2)\}^{\frac{d}{2}}} \, .
\] 
Identity holds if and only if $f$ is elliptical uniform with $\max(f) = M_{f}$.
\end{lemma}

\begin{remark}
When $d=1$, we recover 
$\sigma_{f} \geq 1 / (M_{f} \, \sqrt{12})$, the result obtained
in \cite{ChiaNakano1d}.
\end{remark}
 % \label{lm:minDetCovd}
The proof of the Lemma~\ref{lm:MinDetCovd} has been relegated to
Section~\ref{sec:proofLmMinDetCovd} of the appendix 
to enhance the flow of the paper.
We use the results of Lemma~\ref{lm:MinDetCovd} to prove Theorem~\ref{th:ud}.
\begin{proof}[Proof of Theorem~\ref{th:ud}]
Let $u$ be an elliptical uniform density on 
${\bf x} \in {\mathcal R}^{d} \, (d \geq 1)$. We need to prove that for any set of
mixture components $\{v_{1}, \ldots, v_{m}\}$ of $u$,
\[
|\Sigma_{v_{1}}|^{\frac{1}{2}} + \ldots + |\Sigma_{v_{m}}|^{\frac{1}{2}} 
> |\Sigma_{u}|^{\frac{1}{2}} \,.
\]
Without loss of generality, set $\max(u)=M$ and therefore 
\[
|\Sigma_{u}|^{\frac{1}{2}} = \frac{\Gamma(\frac{d}{2}+1)}
{M \, \{\pi (d+2)\}^{\frac{d}{2}}} \,.
\]
Rewriting the elliptical uniform density $u$ as mixture components, we have
\[
u({\bf x}) = \alpha_{1} \, v_{1}({\bf x}) + \ldots
+ \alpha_{m} \, v_{m}({\bf x})
\]
for some $\{\alpha_{1}, \ldots, \alpha_{m} \}$ satisfying $0 \leq \alpha_{j}
\leq 1$ and $\Sigma \alpha_{j} = 1$. As a result, we have
\[
v_{j}({\bf x}) \leq \frac{u({\bf x})}{\alpha_{j}} \leq \frac{M}{\alpha_{j}} 
\]
for all $1 \leq j \leq m$.
Using Lemma~$\ref{lm:MinDetCovd}$, we have
\begin{equation}
\label{eq:uniformFragments}
|\Sigma_{v_{j}}|^{\frac{1}{2}} \geq 
\frac{\alpha_{j} \, \Gamma(\frac{d}{2}+1)}{M \,\{\pi \, (d+2)\}^{\frac{d}{2}}} = 
\alpha_{j} \, |\Sigma_{u}|^{\frac{1}{2}}  %\\
\end{equation}
for all $j$, with equalities holding if and only if the density in question is 
elliptical uniform. 
Now, for $d>1$, we can have {\it at most} $(m-1)$ but 
{\it never all} of $v$'s to be elliptical uniform satisfying
Eq~(\ref{eq:uniformFragments}).
Therefore,
\[
|\Sigma_{v_{1}}|^{\frac{1}{2}} + \ldots + |\Sigma_{v_{m}}|^{\frac{1}{2}} \geq
|\Sigma_{u}|^{\frac{1}{2}} \,.
\]
Identity may only hold when $d=1$, refer to \cite{ChiaNakano1d}. %\qed
\end{proof}

\subsection{Elliptical Unimodal Densities}
\label{sec:EUD}
In one-dimension, symmetry is trivial to visualize and express mathematically. 
In higher dimensions, symmetry may be depicted via ellipticity. As such, 
{\it elliptical unimodal densities} play a key role in this paper. We provide a 
definition for elliptical unimodal densities below. 
Elliptical densities in general have been treated in detail by many
researchers, see \cite{FangKotzNg} and
references within. Unimodal densities have also been the subject of active
research. For example, refer to \cite{Anderson},
\cite{DharmadhikariBook} as well as \cite{Ibragimov}.

\begin{definition}[Elliptical Unimodal Densities]
%\label{def:ellipticalUniform}
\label{def:ellipticalUnimodal}
We say that $f$ is {\em elliptical unimodal} if there exist a vector $\mu \in
\mathcal{R}^{d}$, a positive semidefinite symmetric matrix 
$\Sigma \in \mathcal{R}^{d \times d}$ and a
{\em non-increasing positive} function $p$ on $\mathcal{R}^{+} \cup \{0\}$ 
such that
\[
f({\bf x}) = p\{({\bf x} - \mu)^{T} \, \Sigma^{-1} \, ({\bf x} - \mu)\} \,.
\]
\end{definition}
Comparing with Definition~\ref{def:spherical}, the only additional information
in Definition~\ref{def:ellipticalUnimodal} is that the positive function $p$ has
to be non-increasing as well.
According to Definition~\ref{def:ellipticalUnimodal}, 
elliptical unimodal densities are those whose
cross-sections are elliptical, and with mean ($\mu$) and covariance
matrices proportional to ($\Sigma$). 
%The class of 
%elliptical unimodal densities form a subset of elliptical densities. 
%(Refer to Definition~\ref{def:spherical}.)
Definition~\ref{def:ellipticalUnimodal} 
encompasses a large class of general densities including $d$-dimensional 
elliptical uniform, Gaussian, logistic, Laplace, Von Mises, beta($k,k$) where
$k>1$, student-$t$, and many other densities.

Henceforth, we propose the following alternative
representation of elliptical unimodal densities.
\begin{theorem}[Representation of Elliptical Unimodal Densities]
\label{th:representationEllipticalUnimodal}
Let $f$ be an elliptical unimodal density with mean $\mu$ and covariance matrix 
$\Sigma$. Then, for all $\epsilon > 0$, 
it is possible to construct a density 
\[
g_{n}({\bf x}) =  \sum_{j=1}^{n} b_{j} \, u_{j}({\bf x})
\]
such that
\[
\int \, |g_{n}({\bf x}) - f({\bf x})| \, d{\bf x} < \epsilon \,.
\]
Here, each $u_{j}$ is an elliptical uniform density such that
\begin{equation}
\label{eq:ellipticalUniformRepresentation}
u_{j}({\bf x}) \propto 
\mathbb{I}_{({\bf x} - \mu)^{T} \, \Sigma^{-1} \, ({\bf x} - \mu) < r_{j}^{2}}
\, 
\end{equation}
and $r$'s are strictly positive. Furthermore, each proportionality constant 
$b_{j}$ satisfies
\[
b_{j} = \frac{r_{j}^{d}}{\sum_{i=1}^{n} r_{i}^{d}} \, .
\]
\end{theorem}
From the above representation, each elliptical uniform component is weighted
proportionally to the hypervolume of its cross-section. The original elliptical
unimodal density is ``sliced lattitudinally" into elliptical uniforms with a
prefixed constant ``thickness". The proof of
Theorem~\ref{th:representationEllipticalUnimodal} has
been relegated to Section~\ref{sec:proofThRepresentation} of the appendix.

\subsection{A Theorem on Elliptical Unimodal Densities}
\begin{theorem}[Inequality on Elliptical Unimodal Densities]
\label{th:maind}
Let $f$ be an elliptical unimodal density with finite second moments. Then, for
any set of mixture components $\{g_{1}, \ldots, g_{m}\}$,
\[
|\Sigma_{f}|^{\frac{1}{2}} \leq |\Sigma_{g_{1}}|^{\frac{1}{2}} + \ldots
+ |\Sigma_{g_{m}}|^{\frac{1}{2}} \, .
\]
Identity is possible only when $f$ is uniform in one-dimension.
\end{theorem}

\begin{proof}
Our task is to prove that for all mixture components 
$\{g_{1}, \ldots, g_{m}\}$
satisfying
\begin{equation}
\label{eq:mixtureComponents}
f({\bf x}) = \sum_{i=1}^{m} \, 
a_{i} \, g_{i}({\bf x}) \,,
\end{equation}
where $0 < a_{i} < 1$ and $\Sigma \, a_{i} = 1$, we must have 
\begin{equation}
|\Sigma_{f}|^{\frac{1}{2}} \leq |\Sigma_{g_{1}}|^{\frac{1}{2}} + \ldots +
|\Sigma_{g_{m}}|^{\frac{1}{2}} \tag{Claim 1} \,.
\end{equation}
%
%Without loss of generality, we may set $f$ to be spherical unimodal. This is
%because we are only interested in the determinant of covariance matrices. 
Using Theorem~\ref{th:representationEllipticalUnimodal}, 
we can approximate $f$ to an arbitrary level of accuracy by rewriting $f$ 
as a finite mixture of elliptical uniform densities, each
having ``uniform thickness" as
\begin{equation}
\label{eq:uniformMixture}
f({\bf x}) = \sum_{j=1}^{n} b_{j} \, u_{j}({\bf x}) \,.
\end{equation}
The ``thickness" of each elliptical uniform component is equal to 
$
\max\{b_{j} u_{j}({\bf x})\} \,.
$
Here, $u_{j}$'s, as described in
Eq~$(\ref{eq:ellipticalUniformRepresentation})$,
are elliptical uniform densities sharing the same
means and whose covariances are multiples of each other.
Each constant of proportionality,
denoted by $b_{j}$, is proportional to the hypervolume of the corresponding 
elliptical uniform density $u_{j}$. 

To provide a link between Eqs~(\ref{eq:mixtureComponents}) and 
(\ref{eq:uniformMixture}), we further rewrite $f$ as
\[
%\begin{equation}
%\label{eq:mesh}
f({\bf x}) = \sum_{i=1}^{m} \sum_{j=1}^{n} c_{i,j} \, v_{i,j}({\bf x}) 
		   = \sum_{i=1}^{m} a_{i} \, g_{i}({\bf x}) 
		   = \sum_{j=1}^{n} b_{j} \, u_{j}({\bf x}) \,.
%\end{equation}
\]
For each pair of $(i,j)$ above, $c_{i,j} \,
v_{i,j}({\bf x})$ is the ``intersection" of the segments 
$a_{i} \, g_{i}({\bf x})$ and $b_{j} \,u_{j}({\bf x})$ with respect to $f$ 
on the curve. 
For all values of $\{i,j\}$, $g_{i}$ and $u_{j}$ can be expressed in terms of 
$v_{i,j}$ as
\begin{equation}
\label{eq:guv}
a_{i} \, g_{i}({\bf x}) = \sum_{j=1}^{n} c_{i,j} \, v_{i,j}({\bf x}) \,, \quad
b_{j} \, u_{j}({\bf x}) = \sum_{i=1}^{m} c_{i,j} \, v_{i,j}({\bf x}) \,.
\end{equation}
Here, depending on the mixture components 
$\{g_{1}, \ldots, g_{m}\}$, it is possible for some of $c_{i,j}$'s to be $0$, 
as long as for all values of $\{i,j\}$, we have
\[
 a_{i} = \sum_{j=1}^{n} c_{i,j} > 0 \,, \quad
 b_{j} = \sum_{i=1}^{m} c_{i,j} > 0 \,.
\]
If $c_{i,j}>0$ for a pair of $(i,j)$, then $v_{i,j}({\bf x})$ is a density.
From Eq~(\ref{eq:guv}), we can rewrite each elliptical uniform $u_{j}$ as
\[
u_{j}({\bf x}) = \sum_{i=1}^{m} \frac{c_{i,j}}{b_{j}} \, v_{i,j}({\bf x}) \,.
\]
Following the argument presented in Theorem~$\ref{th:ud}$, 
%from $\mathcal{M}$-undecomposability of elliptical uniform $u_{j}$, 
we have 
\[
%\begin{equation}
%\label{eq:vbound}
|\Sigma_{v_{i,j}}|^{\frac{1}{2}} \geq \frac{c_{i,j}}{b_{j}} \,
|\Sigma_{u_{j}}|^{\frac{1}{2}} \,,
%\end{equation}
\]
with equality holding if and only if $v_{i,j}$ is elliptical uniform having
``thickness" satisfying
\[
\max\{c_{i,j} \, v_{i,j}({\bf x})\} = \max\{b_{j} \, u_{j}({\bf x})\} \,.
\]
Similarly, rewriting each mixture component $g_{i}$ in terms of $v_{i,j}$, we
obtain
\[
g_{i}({\bf x}) = \sum_{j=1}^{n} \frac{c_{i,j}}{a_{i}} \, v_{i,j}({\bf x}) 
\equiv \sum_{j=1}^{n} s_{i,j} \, v_{i,j}({\bf x}) \,.
\] 
%We have denoted $c_{i,j} / a_{i}$ by $s_{i,j}$ to simplify expressions.

Next, we create new {\it spherical unimodal} densities 
$\tilde{g}_{i}$'s corresponding to
each $g_{i}$ to facilitate lower boundings of $|\Sigma_{g_{i}}|$. Define
$\tilde{g}_{i}$ as follows:
\[
\tilde{g}_{i}({\bf x}) = \sum_{j=1}^{n} \frac{c_{i,j}}{a_{i}} \,
\tilde{v}_{i,j}({\bf x}) 
\equiv \sum_{j=1}^{n} s_{i,j} \, \tilde{v}_{i,j}({\bf x}) \,.
\]
In the above, each $\{\tilde{v}_{i,1}, \ldots, \tilde{v}_{i,n} \}$ 
are {\it spherical uniforms} whose means coincide and such that
\[
\max\{c_{i,j} \, \tilde{v}_{i,j}({\bf x})\} = \max\{b_{j} \, u_{j}({\bf x})\}
\]
for all $\{i,j\}$, hence yielding
\[
%\begin{equation}
%\label{eq:vequal}
|\Sigma_{\tilde{v}_{i,j}}|^{\frac{1}{2}} = \frac{c_{i,j}}{b_{j}} \,
|\Sigma_{u_{j}}|^{\frac{1}{2}} \,.
%\end{equation}
\]
Computing the determinant of the covariance matrix of $g_{i}$, we have
\[
\begin{split}
|\Sigma_{g_{i}}| &= |(s_{i,1} \, \Sigma_{v_{i,1}} + \cdots + s_{i,n} \, \Sigma_{v_{i,n}})
+ (s_{i,1} \, \mu_{v_{i,1}} \mu_{v_{i,1}}^{T} + \cdots 
+ s_{i,n} \, \mu_{v_{i,n}} \mu_{v_{i,n}}^{T})| \\
    &\geq |s_{i,1} \, \Sigma_{v_{i,1}} + \cdots + s_{i,n} \, \Sigma_{v_{i,n}}| \\
        &\geq (s_{i,1} \, |\Sigma_{v_{i,1}}|^{\frac{1}{d}} + \cdots 
          + s_{i,n} \, |\Sigma_{v_{i,n}}|^{\frac{1}{d}})^{d} \\
        &\geq (s_{i,1} \, |\Sigma_{\tilde{v}_{i,1}}|^{\frac{1}{d}} + \cdots 
          + s_{i,n} \, |\Sigma_{\tilde{v}_{i,n}}|^{\frac{1}{d}})^{d} \\          
    &= |s_{i,1} \, \Sigma_{\tilde{v}_{i,1}} + \cdots + s_{i,n} \,
	\Sigma_{\tilde{v}_{i,n}}|
\\
    &=|\Sigma_{\tilde{g}_{i}}|.
\end{split}
\]
The first inequality holds as a result of
\begin{equation}
\label{eq:CT1}
|K_{1} + K_{2}| \geq |K_{1}| \,,
\end{equation}
where $K_{1}$ and $K_{2}$ are both non-negative definite symmetric $d \times d$ 
matrices.
The second inequality holds because
\begin{equation}
\label{eq:CT2}
|K_{1} + K_{2}|^{\frac{1}{d}} \geq 
|K_{1}|^{\frac{1}{d}} + |K_{2}|^{\frac{1}{d}} \,,
\end{equation}
with identity holding if and only if $K_{1}$ and $K_{2}$ are proportional.
The proof of both Eqs~(\ref{eq:CT1}) and (\ref{eq:CT2}) can be found in 
\cite{CoverThomas}.
%The first two inequalities are the direct result of Theorem~$\ref{th:CT1}$ 
%and Theorem~$\ref{th:CT2}$ given in Cover and
%Thomas~[$\ref{cite:CoverThomas}$]. 
The third inequality holds as we must have
\[
|\Sigma_{v_{i,j}}| \geq |\Sigma_{\tilde{v}_{i,j}}| \, 
\]
as a direct result of Lemma~$\ref{lm:MinDetCovd}$. The equality that follows
the third inequality is again a result of Eq~(\ref{eq:CT2}), as all 
$\Sigma_{\tilde{v}_{i,j}}$'s are proportional to the identity matrix.
We have just shown that
\[
|\Sigma_{g_{i}}| \geq |\Sigma_{\tilde{g}_{i}}|
\]
for all $g_{i}$, {\it i.e.} the pseudo-volume of each $g_{i}$ is minimized when
$g_{i}$ is spherical unimodal. %elliptical uniform. 
Therefore, a sufficient condition to (Claim 1) is
\begin{equation}
|\Sigma_{f}|^{\frac{1}{2}} \leq |\Sigma_{\tilde{g}_{1}}|^{\frac{1}{2}} 
+ \ldots +|\Sigma_{\tilde{g}_{m}}|^{\frac{1}{2}} \tag{Claim 2} \,.
\end{equation} 

Since $f$ is elliptical unimodal, it is possible to find a corresponding
spherical unimodal density $f^{s}$ such that the hypervolumes are preserved,
{\it i.e.} $|f^{s}| = |f|$.
To prove (Claim 2), we only have to deal with the pseudo-volumes of 
spherical %elliptical
unimodal densities.
We obtain $|\Sigma_{\tilde{g_{i}}}|$ as follows
\[
|\Sigma_{\tilde{g_{i}}}| 
%=\frac{1}{(d+2)^{d}} \cdot 
%(\frac{s_{i,1}^{1+\frac{2}{d}}+\cdots+s_{i,n}^{1+\frac{2}{d}}} 
%{s_{i,1}+\cdots+s_{i,n}})^{d}
=\frac{1}{(d+2)^{d}} \cdot 
(\frac{c_{i,1}^{1+\frac{2}{d}}+\cdots+c_{i,n}^{1+\frac{2}{d}}} 
{c_{i,1}+\cdots+c_{i,n}})^{d} \,.
\]
Here, we make use of the fact that the covariance of a
$d$-dimensional spherical uniform %unimodal 
density defined by
\[
u({\bf x}) \propto \mathbb{I}_{|{\bf x} - \mu| < r}
\]
 is given as
\[
\Sigma_{u} = \frac{r^{2}}{(d+2)} \cdot {\mathbf I}_{d} \,
\]
where ${\mathbf I}_{d}$ denotes the identity matrix in $d$-dimensional space.
Refer to Eq~\eqref{eq:covOfSphericalUniform}. Similarly, 
\[
|\Sigma_{f}| =
\frac{1}{(d+2)^{d}} \cdot 
(\frac{b_{1}^{1+\frac{2}{d}}+\cdots+b_{n}^{1+\frac{2}{d}}} 
{b_{1}+\cdots+b_{n}})^{d} \,.
\]
Hence, proving (Claim 2) is equivalent to proving 
\begin{equation}
(\frac{b_{1}^{1+\frac{2}{d}} + \ldots + b_{n}^{1+\frac{2}{d}}}
{b_{1}+\ldots+b_{n}})^{\frac{d}{2}}
\leq
(\frac{c_{1,1}^{1+\frac{2}{d}} + \ldots + c_{1,n}^{1+\frac{2}{d}}}
{c_{1,1}+\ldots+c_{1,n}})^{\frac{d}{2}}
+ \ldots +
(\frac{c_{m,1}^{1+\frac{2}{d}} + \ldots + c_{m,n}^{1+\frac{2}{d}}}
{c_{m,1}+\ldots+c_{m,n}})^{\frac{d}{2}} \,, \tag{Claim 3}
\end{equation}
where $b_{j}= c_{1,j} + \ldots + c_{m,j}$ for all $j$. To prove (Claim 3), 
we just have to invoke Lemma~$\ref{lm:ineqIntermediated}$ given below for a
total of ($m-1$) times, adding up
summands on the RHS two at a time and maintaining the ``$\leq$" sign.
We are now left with proof of Lemma~$\ref{lm:ineqIntermediated}$ to prove
Theorem~$\ref{th:maind}$.
\renewcommand{\qedsymbol}{}
\end{proof}

%\begin{inequality}
\begin{lemma}
\label{lm:ineqIntermediated}
Let $a_{i}, b_{i}, c_{i}$ be sequences of non-negative real numbers such that 
for all $i$, $a_{i}=b_{i}+c_{i}$ and $a_{i}>0$. Then the following
inequality holds for any positive integers $d$ and $n$. 
%$d, n \in \mathcal{N}^{+}$:
\[
(\frac{a_{1}^{1+\frac{2}{d}}+\cdots+a_{n}^{1+\frac{2}{d}}} 
{a_{1}+\cdots+a_{n}})^{\frac{d}{2}} 
\leq
(\frac{b_{1}^{1+\frac{2}{d}}+\cdots+b_{n}^{1+\frac{2}{d}}} 
{b_{1}+\cdots+b_{n}})^{\frac{d}{2}}
+
(\frac{c_{1}^{1+\frac{2}{d}}+\cdots+c_{n}^{1+\frac{2}{d}}} 
{c_{1}+\cdots+c_{n}})^{\frac{d}{2}} \, .
\]
Equality holds if and only if the sequences $a_{i}, b_{i}$ and $c_{i}$ are
linearly dependent.
%\end{inequality}
\end{lemma}

\begin{proof}
The proof is similar to that of \cite{ChiaNakano1d}, with the only
difference being in $d$. We proceed in the spirit of
\cite{HardyLittlewoodPolya}, as well as \cite{PolyaSzego}.
Set $\mathbf{x} \equiv [x_{1},\cdots,x_{n}]^{T}$, $\mathbf{y} \equiv 
[y_{1},\cdots,y_{n}]^{T}$ and $\mathbf{z} \equiv [z_{1},\cdots,z_{n}]^{T}$ 
and similarly for $\mathbf{a, b, c}$.
%\newline
Let $\mathbf{x} = t \, \mathbf{y} + (1-t) \, \mathbf{z}$, {\it i.e.} 
$x_{i}=t\,y_{i}+(1-t)\,z_{i}$ for all $i$. 
Furthermore, define the function $f$ as follows:
\[f(\mathbf{x})=(\frac{x_{1}^{1+\frac{2}{d}}+\cdots+x_{n}^{1+\frac{2}{d}}} 
{x_{1}+\cdots+x_{n}})^{\frac{d}{2}} \,, 
\]
and set
\[
\phi(t) = f\{t \, \mathbf{y} + (1-t) \, \mathbf{z}\} \equiv f(\mathbf{x}) \,,
\]
where $0 \leq t \leq 1$. %To prove Lemma~$\ref{lm:ineqIntermediated}$, 
It suffices to prove that $\phi^{\prime\prime}(t) \geq 0$ for $0 \leq t \leq 1$.
This is an immediate
consequence of Jensen's inequality as $\phi^{\prime\prime}(t)\geq0$ implies 
\[
\phi(t) \leq t \, \phi(0) + (1-t) \, \phi(1) \,.
\]
Setting $t=\frac{1}{2}$ \,, 
we have
\[
f(\frac{\mathbf{y}}{2}+\frac{\mathbf{z}}{2})\leq\frac{1}{2}\,f(\mathbf{y})
+\frac{1}{2}\,f(\mathbf{z}) \,.
\]
Denoting by $\mathbf{y=b},\, \mathbf{z=c}$, this becomes 
\[
f(\frac{\mathbf{a}}{2})\leq \frac{1}{2} \, f(\mathbf{b}) + \frac{1}{2} \, f(\mathbf{c}) \,.
\]
However, from the definition of $f$, we must have 
\[
f(\frac{\mathbf{a}}{2}) 
%= (\frac{1}{2})^{(1+\frac{2}{d}-1) \cdot \frac{d}{2}} \, f(\mathbf{a}) 
= \frac{1}{2} f(\mathbf{a}) \,.
\] 
%or $f(\frac{\mathbf{a}}{2}) = \frac{1}{2} \cdot f(\mathbf{a}).$ 
Therefore $\phi^{\prime\prime}(t) \geq 0$ implies 
$f(\mathbf{a}) \leq f(\mathbf{b}) + f(\mathbf{c})$ as required. 
Equality holds if and only if $\phi^{\prime\prime}(t) = 0$.

We shall begin from the definition of $\phi$ as follows:
\[
%\begin{equation}
%\label{eq:phi_t0d}
\phi(t) = f(\mathbf{x}) = (\Sigma \, x_{i}^{1+\frac{2}{d}})^{\frac{d}{2}} \, 
					(\Sigma \, x_{j})^{-\frac{d}{2}}.
%\end{equation}
\]
Differentiating $\phi$ twice with respect to $t$ and rearranging, we have
\[
%\begin{equation}
\begin{split}
%\label{eq:phi/phid}
\frac{\phi^{\prime\prime}(t)}{\phi(t)}
&= \frac{d \, (d+2)}{4} \cdot \underbrace{\{\,%[\Sigma x_{j}]^{-1} \cdot 
\frac{\Sigma (y_{k}-z_{k})}{\Sigma x_{j}} - %[\Sigma x_{i}^{1+\frac{2}{d}}]^{-1} \cdot 
\frac{\Sigma x_{k}^{\frac{2}{d}} \, (y_{k} - z_{k})}
{\Sigma x_{i}^{1+\frac{2}{d}}} \,\}^{2}}_{A} \\
&+ (\frac{d+2}{d})\cdot (\Sigma x_{i}^{1+\frac{2}{d}})^{-2}
 \cdot \underbrace{[\, (\Sigma x_{i}^{1+\frac{2}{d}})
\cdot \{\Sigma x_{j}^{\frac{2}{d}-1} (y_{j}-z_{j})^{2}\} 
- \{\Sigma x_{k}^{\frac{2}{d}}\,(y_{k}-z_{k})\}^{2} \,]}_{B} \,.
\end{split}
%\end{equation}
\]
The term $A$ is expressible as a square and therefore greater or equal to $0$.
To evaluate $B$, we set $p_{i}^{2} = x_{i}^{1+\frac{2}{d}}$ and
$q_{j}^{2}=x_{j}^{\frac{2}{d}-1} (y_{j}-z_{j})^{2}$, yielding
\[
%\begin{equation}
B = (\Sigma p_{i}^{2}) \cdot (\Sigma q_{j}^{2}) - (\Sigma p_{k}\,q_{k})^{2} 
\geq 0 \,
%\end{equation}
\]
via Cauchy-Schwarz's inequality. 
Therefore we must have
\[\phi^{\prime\prime}(t) \geq 0\] due to the non-negativeness of $x_{i},y_{i}$
and $z_{i}$. Hence, Lemma~\ref{lm:ineqIntermediated}, 
and consequently, Theorem~\ref{th:maind}
is proved. %\qed
\end{proof}

As a result of Theorem~\ref{th:maind}, all elliptical unimodal
densities with finite second moments are $\mathcal{M}$-undecomposable.
Conversely, any density, which is  
$\mathcal{M}$-decomposable, cannot be elliptical unimodal. 
One can do better than that.
In the next subsection, we further show that 
if $f$ is $\mathcal{M}$-decomposable, then there exists an approximation
to represent $f$ via a mixture of Gaussian densities, which improves 
estimation of $f$.
\subsection{Estimation of $\mathcal{M}$-Decomposable Densities}
\begin{theorem}[Inequality on $\mathcal{M}$-Decomposable Densities]
% and Kullback-Leibler Divergence]
\label{th:KLD}
Let $f$ be probability density functions defined on ${\bf x} \in {\mathcal
R}^{d}$. Let $\{g_{1}, \ldots ,g_{m}\}$ be a set of mixture components of $f$ 
such that 
\[
f({\bf x}) 
 = \alpha_{1} \, g_{1}({\bf x}) + \ldots + \alpha_{m} \, g_{m}({\bf x}) \,,
\] where $0 < \alpha_{j} < 1$ and $\Sigma \alpha_{j} = 1$.
Then the following result applies:
\[
\begin{split}
	%\text{\em If \ } 
	|\Sigma_{f}|^{\frac{1}{2}} &> 
	|\Sigma_{g_{1}}|^{\frac{1}{2}} + \ldots + |\Sigma_{g_{m}}|^{\frac{1}{2}} \\ 
	\Rightarrow \quad
	%\text{\em then \ } 
	KL(f \, \| \, \tilde{f}) &> KL(f \, \| \, \alpha_{1} \, \tilde{g}_{1} 
	+ \ldots + \alpha_{m} \, \tilde{g}_{m}) \,.
\end{split}
\]
Here, $KL(\, p \, \| \, q \,)$ denotes the Kullback-Leibler divergence between
densities $p$ and $q$, given as
\[
KL(\, p \, \| \, q \, ) = \int p({\bf x}) \log \frac{p({\bf x})}{q({\bf x})} \, 
d{\bf x} \,.
\] 
Furthermore, $\tilde{f}$ denotes the Gaussian density
%with $\mu_{\tilde{f}} = \mu_{f}$, $\Sigma_{\tilde{f}} = \Sigma_{f}$; while
whose mean and covariance matrix coincide with those of $f$, and 
$\tilde{g}$'s are similarly defined.

\end{theorem}

\begin{proof}
We only need to prove that
 \begin{equation}
 %\label{eq:KLD}
 \int f(\mathbf{x}) \, \log \tilde{f} (\mathbf{x}) \, d\mathbf{x} < 
 \int f(\mathbf{x}) \, \log \{\alpha_{1} \, \tilde{g}_{1} (\mathbf{x}) 
 + \ldots + \alpha_{m} \, \tilde{g}_{m} (\mathbf{x})\} \, d\mathbf{x} \,.   
 \tag{Claim A}
 \end{equation}
Now, RHS of (Claim A) %Eq~(\ref{eq:KLD})
\begin{equation*}
\begin{split}
 &= \int \{\alpha_{1} \, g_{1}(\mathbf{x}) 
 + \ldots + \alpha_{m} \, g_{m}(\mathbf{x})\}  \cdot 
     \log \{\alpha_{1} \, \tilde{g}_{1}(\mathbf{x}) + \ldots + \alpha_{m} \,
\tilde{g}_{m} (\mathbf{x}) \} \, d\mathbf{x} \\
    &\geq \alpha_{1} \int  g_{1}(\mathbf{x}) \,
     \log \{\alpha_{1} \, \tilde{g}_{1} (\mathbf{x})\} \, d\mathbf{x} + \ldots 
     + \alpha_{m} \int  g_{m}(\mathbf{x}) \,
     \log \{\alpha_{m} \, \tilde{g}_{m} (\mathbf{x})\} \, d\mathbf{x} \\
     &= \alpha_{1} \, \{\log \alpha_{1} + \int g_{1}(\mathbf{x}) \log
	 \tilde{g}_{1} (\mathbf{x}) \, d\mathbf{x} \, \} + \ldots %\\
	 + \alpha_{m} \, \{\log \alpha_{m} + \int g_{m}(\mathbf{x}) 
	 \log \tilde{g}_{m} (\mathbf{x}) \, d\mathbf{x} \,\}.
\end{split}
\end{equation*}
From definitions, the probabilitiy density function of $\tilde{g}(\mathbf{x})$ 
is given by
\[
\tilde{g}(\mathbf{x}) = (2 \pi)^{-\frac{d}{2}} \, |\Sigma_{g}|^{-\frac{1}{2}} 
\exp \{-\frac{1}{2} \, (\mathbf{x-\mu_{g}})^{T} \, \Sigma_{g}^{-1}
(\mathbf{x-\mu_{g}}) \} \,,
\]
where $\mu_{g}$ and $\Sigma_{g}$ denote the mean and covariance matrix of $g$.
We obtain
\[
\int \, g(\mathbf{x}) \, \log \tilde{g}(\mathbf{x}) \, d\mathbf{x} 
 = {-\frac{d}{2}} \log(2 \pi)  - {\frac{1}{2}} \, \log |\Sigma_{g}| - \frac{d}{2}
\,.
\]
Hence, RHS of (Claim A) % Eq~(\ref{eq:KLD})
\[
\geq	\alpha_{1} \, \{\, \log \alpha_{1} - \frac{1}{2} \log |\Sigma_{g_{1}}| \,\} 
    + \ldots + \alpha_{m} \, \{\, \log \alpha_{m} 
	- \frac{1}{2} \log |\Sigma_{g_{m}}| \,\}
     - \frac{d}{2} \log (2 \pi) - \frac{d}{2}.
\]
Meanwhile, %applying similarly to $f$, we have
\[
\text{LHS of (Claim A)} = {-\frac{d}{2}} \log(2 \pi)  - {\frac{1}{2}} \,
\log |\Sigma_{f}|  - \frac{d}{2}.
\]
To complete the prove of Theorem~\ref{th:KLD}, it
suffices to demonstrate that 
 \begin{equation}
 \alpha_{1} \, \log \frac{|\Sigma_{g_{1}}|^{\frac{1}{2}}}{\alpha_{1}} +
 \ldots + \alpha_{m} \, \log \frac{|\Sigma_{g_{m}}|^{\frac{1}{2}}}
 {\alpha_{m}} 
       <  \log |\Sigma_{f}|^{\frac{1}{2}}.
    \tag{Claim B}
 \end{equation}
 Using Jensen's inequality, we have
 \[
 \begin{split}
 \text{LHS of (Claim B)} 
 &\leq \log ( \alpha_{1} \frac{|\Sigma_{g_{1}}|^{\frac{1}{2}}}{\alpha_{1}}
+ \ldots + \alpha_{m} \frac{|\Sigma_{g_{m}}|^{\frac{1}{2}}}{\alpha_{m}} ) \\
  &= \log (|\Sigma_{g_{1}}|^{\frac{1}{2}} + \ldots + |\Sigma_{g_{m}}|^{\frac{1}{2}}) \\
  &< \log |\Sigma_{f}|^{\frac{1}{2}} = \text{RHS of (Claim B)}, %\qed
 \end{split}
 \]
 which completes the proof of Theorem~$\ref{th:KLD}$ %\qed
\end{proof}

We summarize the result of Theorem~\ref{th:KLD} as follows. 
Let $f$ be any density in
$d$-dimensional space. If $f$ is $\mathcal{M}$-decomposable, then by definition,
% If 
one can find a set of mixture components
of $f$, such that the sum of pseudo-volumes of the mixture components 
is less than the pseudo-volume of the original density $f$.
%then, by definition, $f$ is $\mathcal{M}$-decomposable. From
From Theorem~\ref{th:maind}, 
$f$ cannot belong to the class of elliptical unimodal
densities. It is possible to do better than that. 
Theorem~\ref{th:KLD} shows that
$f$ is better estimated via a weighted Gaussian mixture, rather than a single
Gaussian density. The Gaussian components are created via
moments matching of the mixture components of $f$. The better goodness of fit by
the resultant weighted Gaussian mixture estimate is guaranteed in
Kullback-Leibler sense. It should be noted that the analytical form of the
original density $f$ does not need to be known. In the next section, we
demonstrate the use of Theorems~\ref{th:maind} and \ref{th:KLD} to satistical
applications, namely cluster analysis and kernel density estimation.

\section{Applications Using $\mathcal{M}$-Decomposability}
\label{sec:applications}

\subsection{Clustering via $\mathcal{M}$-Decomposability: The Power of Two}
\label{sec:clustering}

\begin{figure}%[t]
\centerline{\includegraphics[width=100mm,clip]{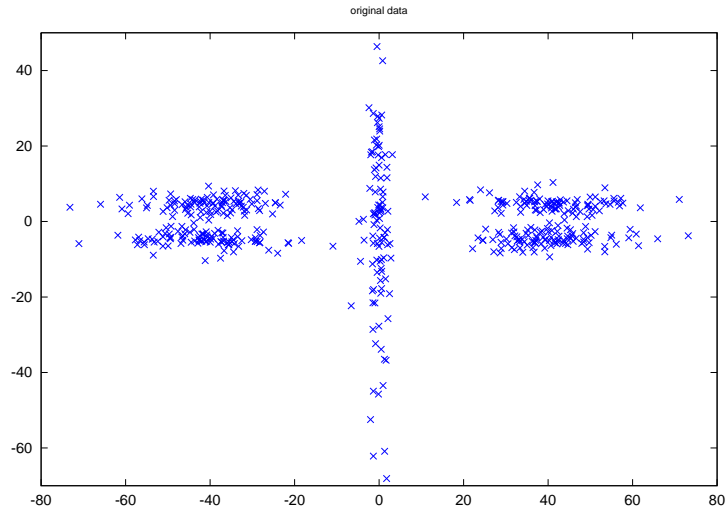}}\vskip2mm
\caption{Original data from multimodal density drawn from mixture of five
logistic densities.}
\end{figure}

\begin{figure}%[p]
\centerline{\includegraphics[width=100mm,clip]{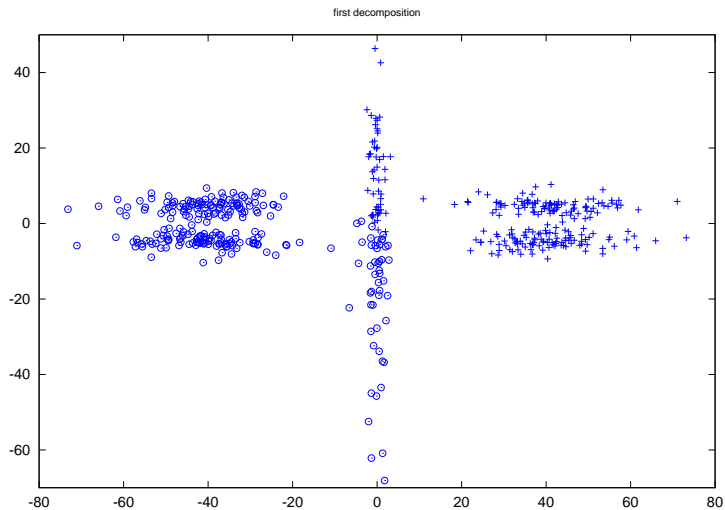}}\vskip2mm
%\caption{Decomposition pair of original density.}
\caption{Original data split into two mixture components, 
represented by two different symbols. 
The sum of pseudo-volumes of the mixture components is less than that of original.}
\end{figure}

One straightforward application of $\mathcal{M}$-decomposability is cluster
analysis. Many existing clustering algorithm divide the dataset into
clusters, based on the following heuristic:
That the within-variances of clusters are minimized while the 
between-variance is maximized at the same time. Another variation to this
heuristic is to determine cluster allocations such that
 a function of volume of clusters is minimized. In particular,
\cite{ShiodaTuncel} proposed dividing the dataset into $k$ clusters, such
that the total sum of MVE (minimum volume of ellipsoid) of $k$ clusters are
globally minimized. While the details for each algorithm may differ, 
the underlying idea is conceptually similar. Theorem~\ref{th:KLD} provides 
theoretical justification for minimizing sum of pseudo-volumes, 
and therefore supports all similar approaches of existing algorithms.

Intuitively, the rigorous approach to implement cluster analysis via
Theorem~\ref{th:KLD} is to divide the dataset into $k (\, \geq 1)$ clusters, 
such that the sum of pseudo-volumes of all clusters are globally minimized.
This approach is computationally unfeasible for dataset of any reasonable
size. To this end, we propose the following alternative approach that captures 
the essence of Theorem~\ref{th:KLD} as far as possible. 
We devise a split-merge clustering strategy that involves splitting and
merging, two clusters at a time. This lowers the overall computational load. 
We show that with our approach, the algorithm is able to overcome local minima. 
Consequently, it is possible to perform cluster analysis well, 
even with $k (\, > 2)$ clusters.

\begin{figure}%[t]
\centerline{\includegraphics[width=100mm,clip]{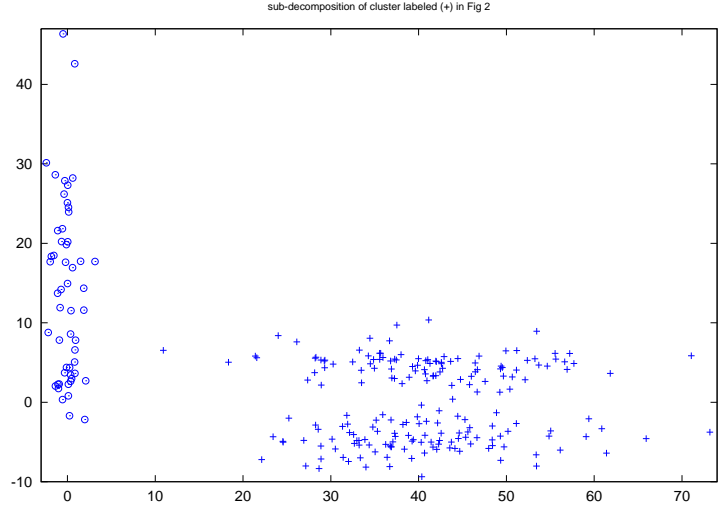}}\vskip2mm
%\caption{Decomposition pair of first mixture component from Fig $2$.}
\caption{Mixture component denoted by (+) in Fig 2 split into two further 
mixture components.}
\end{figure}

\begin{figure}%[t]
\centerline{\includegraphics[width=100mm,clip]{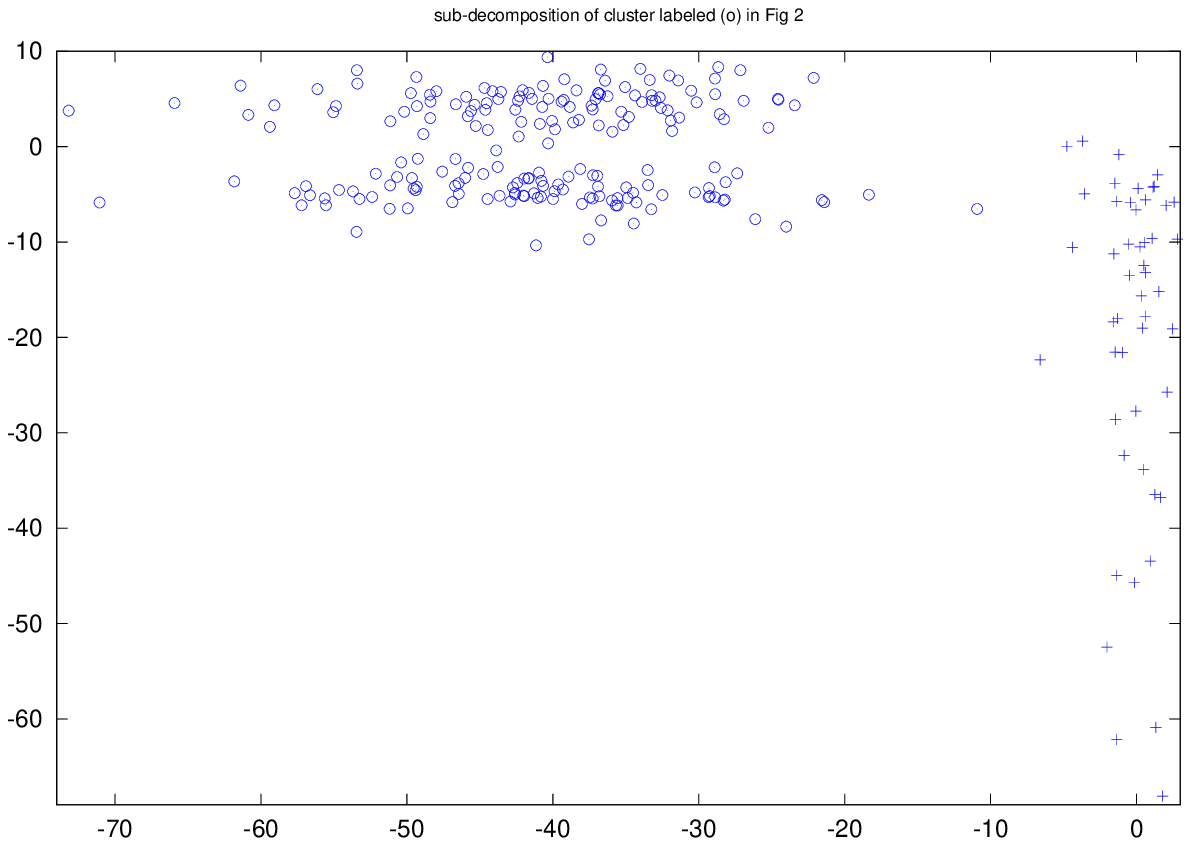}}\vskip2mm
%\caption{Decomposition pair of second mixture component from Fig $2$.}
\caption{Mixture component denoted by (o) in Fig 2 split into two further 
mixture components.}
\end{figure}

\begin{figure}%[t]
\centerline{\includegraphics[width=100mm,clip]{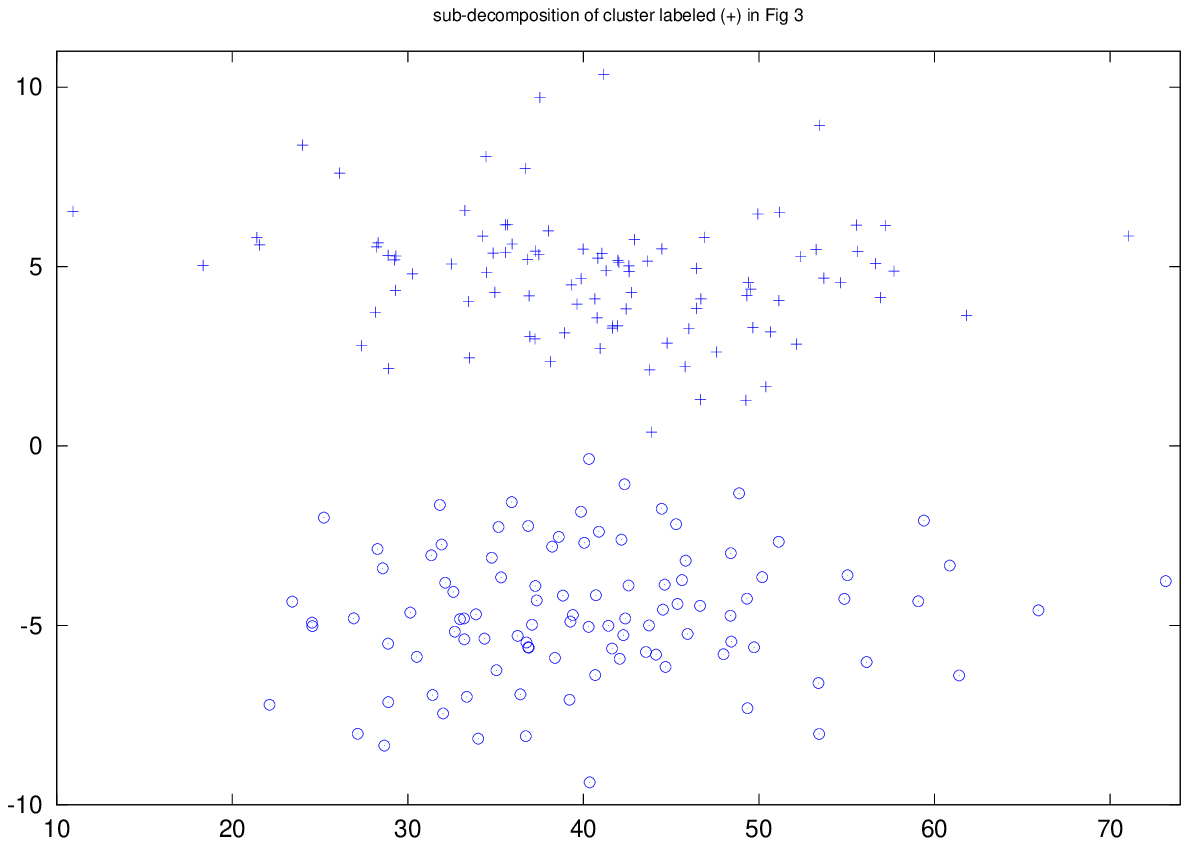}}\vskip2mm
%\caption{Decomposition pair of first mixture component from Fig $2$.}
\caption{Mixture component denoted by (+) in Fig 3 split into two further 
mixture components.}
\end{figure}

\begin{figure}%[t]
\centerline{\includegraphics[width=100mm,clip]{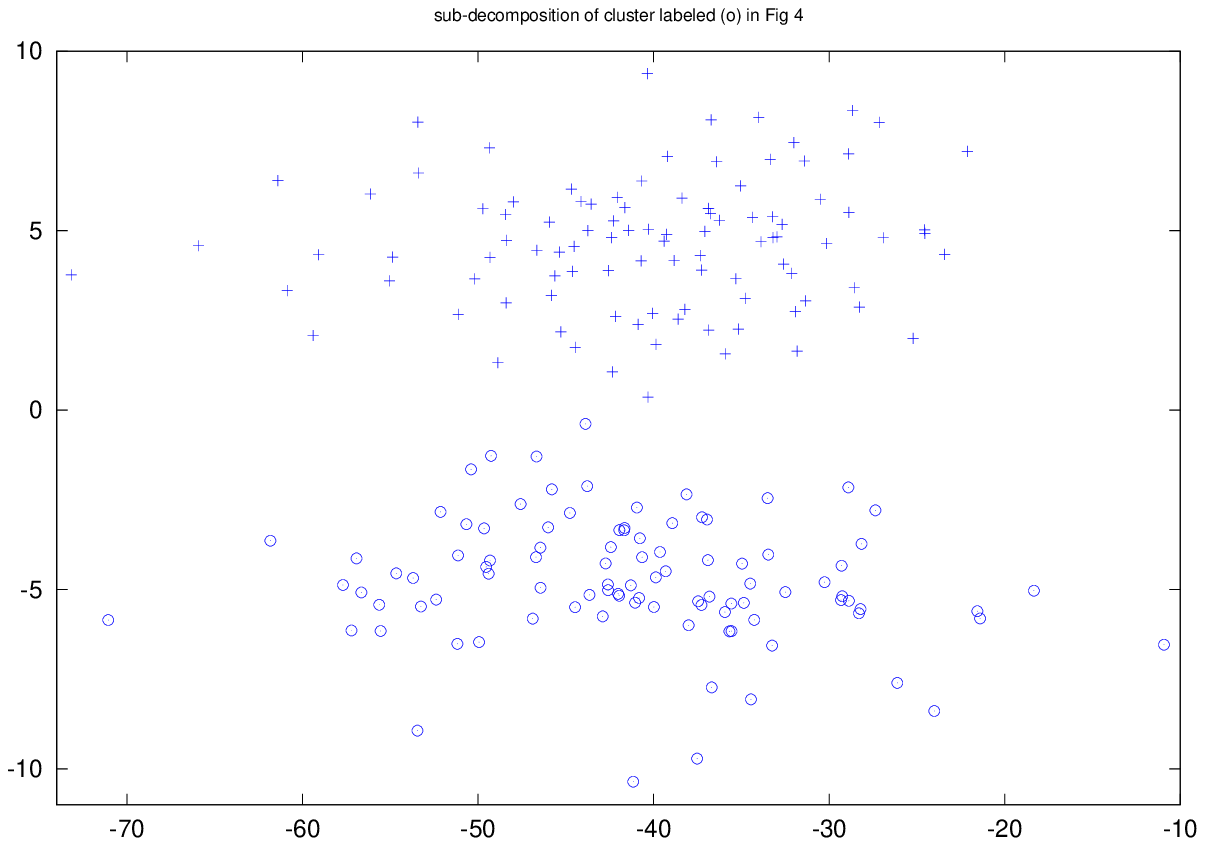}}\vskip2mm
%\caption{Decomposition pair of second mixture component from Fig $2$.}
\caption{Mixture component denoted by (o) in Fig 4 split into two further 
mixture components.}
\end{figure}

From the given sample 
$F = \{X_{1}, \cdots, X_{n}\}$, we are interested to
know if the original sample is $\mathcal{M}$-decomposable. 
We check if $F$ can be partitioned into two clusters, such that the sum
of pseudo-volumes of the clusters is less than that of $F$. 
We denote as $\{G, H\}$, a partition of $F$, such that
\[
G=\{Y_{1}, \cdots, Y_{m}\}, \quad H=\{Y_{m+1}, \cdots, Y_{n}\}
\]
and $G \cup H = F$, with $Y$'s being a rearrangement of $X$. 
We further denote the sample covariance matrices of $F, G, H$ as 
$\mathcal{S}_{F}$, $\mathcal{S}_{G}$ and $\mathcal{S}_{H}$. 
Our task is to find the optimal partition $\{G,H\}$ such that 
\[
|\mathcal{S}_{G}|^{\frac{1}{2}} + |\mathcal{S}_{H}|^{\frac{1}{2}}
\]
is globally minimized and test this value against 
$|\mathcal{S}_{F}|^{\frac{1}{2}}$. 
If
\begin{equation}
\label{eq:MDtestS}
\frac{|\mathcal{S}_{G}|^{\frac{1}{2}} + |\mathcal{S}_{H}|^{\frac{1}{2}} }{
 |\mathcal{S}_{F}|^{\frac{1}{2}} } < 1 + \tau_{s} \,,
\end{equation}
where $\tau_{s}$ is a threshold value close to zero,
then, we can conclude that $F$ is likely to be $\mathcal{M}$-decomposable.
However, if 
Eq~(\ref{eq:MDtestS}) is not satisfied,
then $F$ is likely to be $\mathcal{M}$-undecomposable. 
To robustify the ``splitting process" against local minima traps, it is possible
to set the RHS of Eq~(\ref{eq:MDtestS}) to be greater than $1$. Furthermore, 
taking into consideration error due to finiteness of sample sizes, 
imperfection of splitting algorithms, and also accounting for limiting the
number of mixture components to two, 
we recommend that the $\tau_{s}$ on the RHS of Eq~(\ref{eq:MDtestS}) 
to be about $0.05$.

When one concludes that a particular cluster $F$ is probably 
$\mathcal{M}$-undecom-posable, it is possible to stop at one cluster.
However, if $F$ is found to be $\mathcal{M}$-decomposable into clusters of 
$G$ and $H$, one may repeat the splitting process for $G$ and $H$. 
The process is then reiterated until all clusters are probably 
$\mathcal{M}$-undecomposable. 
When that happens, the splitting process ends.

Our strategy also includes ``merging" of clusters. 
At the point when all splitted clusters are probably 
$\mathcal{M}$-undecomposable, we select two clusters at a time and perform 
the following test.
Now, let $Q, R$ denote the two chosen clusters and $P$ be the union of the
two clusters, {\it i.e.} $P = Q \cup R$. We then check the sum of the
pseudo-volumes of $Q$ and $R$ and compare against that of $P$. If 
\begin{equation}
\label{eq:MDtestM}
\frac{|\mathcal{S}_{Q}|^{\frac{1}{2}} + |\mathcal{S}_{R}|^{\frac{1}{2}} }
{ |\mathcal{S}_{P}|^{\frac{1}{2}} } \geq 1 + \tau_{m} \,,
\end{equation}
we conclude that $Q$ and $R$ should be merged to form a larger cluster $P$.
This process is repeated until there are no more mergeable clusters left. To
prevent overclustering, we recommend $\tau_{m}$ to be around $-0.05$.

We have described a possible algorithm using $\mathcal{M}$-decomposability 
to perform cluster analysis. The crucial point is to find 
a partition $\{G,H\}$ such that  
$|\mathcal{S}_{G}|^{\frac{1}{2}} + |\mathcal{S}_{H}|^{\frac{1}{2}}$ is
minimized as far as possible. 
There are many possible approaches to this task. 
To find the global minimum of the sum
$|\mathcal{S}_{G}|^{\frac{1}{2}} + |\mathcal{S}_{H}|^{\frac{1}{2}}$
 is computationally unfeasible and may be NP-hard. 
Here, we propose a computationally simpler approach. 
At each spitting step, we simply fit a two-mixture Gaussian to the original 
cluster $F$, and then run the EM algorithm to convergence to obtain the 
partition $\{G,H\}$.
However, we emphasize that the EM algorithm approach itself is not critical, 
and that it is possible to use other approaches to obtain a reasonable
partition $\{G,H\}$ of $F$ at the splitting step. The main point here
is the concept of clustering via $\mathcal{M}$-decomposability.  
In the two examples presented below, we show that it is possible to perform
clustering analysis reasonably well, using our proposed algorithm. 

\begin{figure}%[t]
\centerline{\includegraphics[width=100mm,clip]{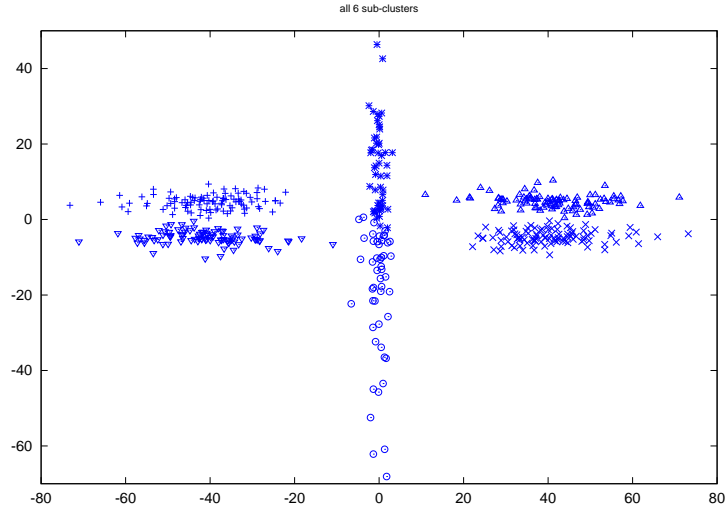}}\vskip2mm
\caption{All six $\mathcal{M}$-undecomposable clusters of original data,
represented by six different symbols.}
\end{figure}

\begin{figure}%[t]
\centerline{\includegraphics[width=100mm,clip]{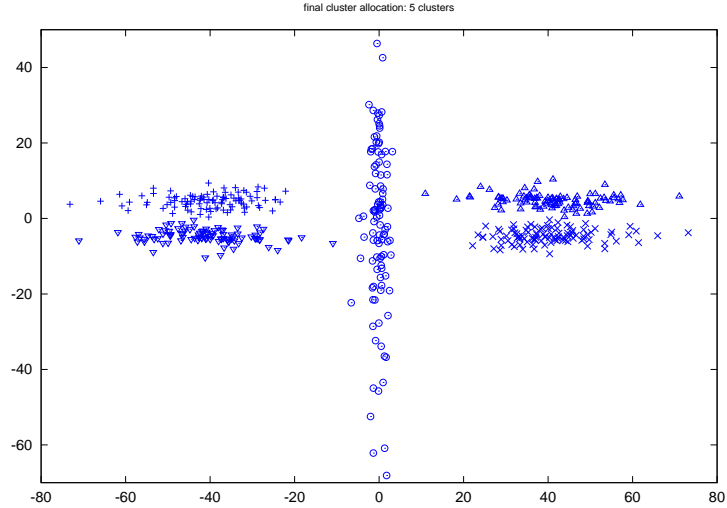}}\vskip2mm
%\centerline{\includegraphics[width=100mm,clip]{figEPS/k3means.eps}}\vskip2mm
\caption{Final cluster allocation formed by merging clusters from Fig 7. Five
 clusters are recovered faithfully.}
\end{figure}

%\begin{figure}%[t]
%\centerline{\includegraphics[width=80mm,clip]{figEPS/k5means_500.eps}}\vskip2mm
%\centerline{\includegraphics[width=100mm,clip]{figEPS/k3means.eps}}\vskip2mm
%\caption{Final cluster allocation formed by merging clusters from Fig 5. Three clusters are recovered
%faithfully.}
%\end{figure}

\subsection{Clustering of Simulated Data}
The simulation example provided here is drawn from a five-mixture logistic
densities as follows. The sample $F$ is generated by
$100$ samples each from five logistic densities with the following means
and covariance matrices:
\[
L_{1} :
\left[
\left( 
\begin{array}{c}
-40 \\
5
\end{array}
\right)
\,,  
\left(
\begin{array}{cc}
12 \, \pi^{2} & 0  \\
0 & \frac{\pi^{2}}{3}
\end{array}
\right)
\right]
\]
\[
L_{2} :
\left[
\left( 
\begin{array}{c}
-40 \\
-5
\end{array}
\right)
\,,  
\left(
\begin{array}{cc}
12 \, \pi^{2} & 0  \\
0 & \frac{\pi^{2}}{3}
\end{array}
\right)
\right]
\]
\[
L_{3} :
\left[
\left( 
\begin{array}{c}
0 \\
0
\end{array}
\right)
\,,  
\left(
\begin{array}{cc}
\frac{\pi^{2}}{3} & 0  \\
0 & 48 \, \pi^{2}
\end{array}
\right)
\right]
\]
\[
L_{4} :
\left[
\left( 
\begin{array}{c}
40 \\
5
\end{array}
\right)
\,,  
\left(
\begin{array}{cc}
12 \, \pi^{2} & 0  \\
0 & \frac{\pi^{2}}{3}
\end{array}
\right)
\right]
\]
\[
L_{5} :
\left[
\left( 
\begin{array}{c}
40 \\
-5
\end{array}
\right)
\,,  
\left(
\begin{array}{cc}
12 \, \pi^{2} & 0  \\
0 & \frac{\pi^{2}}{3}
\end{array}
\right)
\right]
\]
Fig~$1$ shows the original sample $F$.
Clustering is performed without knowledge of either the number of clusters 
or the functional form of the clusters. At the first split step, 
we fit a two-Gassian mixture to $F$, and perform EM to
obtain the partition $\{G,H\}$. 
The result is shown in Fig~$2$. 
As Eq~(\ref{eq:MDtestS}) is satisfied for $F, G, H$, we split
$F$ into $G$ and $H$. 
This is a case of EM converging to a local minima as it
is (visually) unlikely that $G$ and $H$ are meaningful clusters of $F$. 
However, from
Eq~(\ref{eq:MDtestS}), it is theoretically better off to split $F$ into $G$ and
$H$. The theoretical justification is given in Kullback-Leibler sense. 
The splitting process is
repeated for $G$ and $H$ and the results are shown in Figs~$3$ and $4$. 
The splitting process continues until we arrive at six clusters that are
%At this point, all four clusters 
are all $\mathcal{M}$-undecomposable (Fig~$7$). 
Finally, we begin the merging process and find that the two clusters 
$Q$, shown as asterix (*) and $R$, shown as circle (o) in Fig~$7$, satisfy
Eq~(\ref{eq:MDtestS})
where $P=Q \cup R$. The two clusters are then merged and we are left with five
clusters shown in Fig~$8$. This example shows that our algorithm is easy to
implement and is robust to local minima.

\begin{figure}%[t]
%\centerline{\includegraphics[width=100mm,clip]{figEPS/FinalClusters.eps}}\vskip2mm
\centerline{\includegraphics[width=100mm,clip]{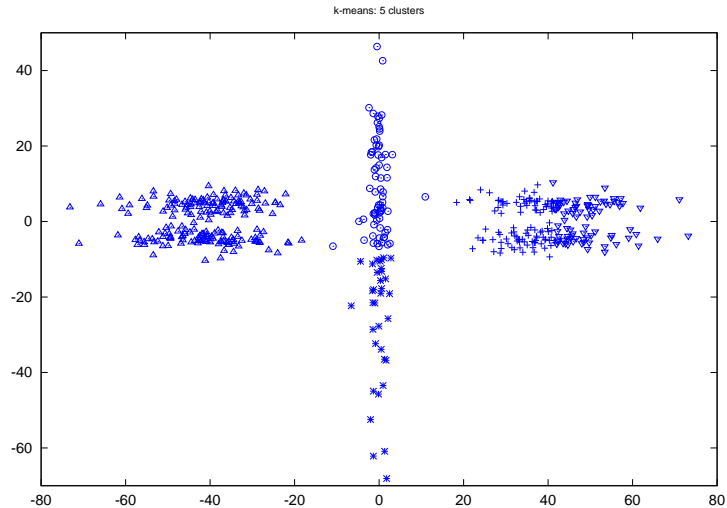}}\vskip2mm
\caption{Final cluster allocation via $k$-means, represented by five different 
symbols. $k$-means algorithm fails to recover clusters faithfully.}
\end{figure}

A popular clustering algorithm is the $k$-means method, which is optimal for
nearly spherical clusters. However, it does not work here because of the 
presence of inherently elongated clusters.
%vertical and horizontal axes are different in
%scale, causing each cluster to be elongated. 
Even by setting $k=5$, the $k$-means method does not achieve a meaningful 
cluster allocation, as shown in Fig~$9$. Cluster analysis via $k$-means is
sensitive to rescaling of axes, because $k$-means involves comparison of
distances. To improve the performance of $k$-means
analysis, there exist many pre-processing heuristics, 
{\it e.g.} rescaling the axes such that all axial units or marginal standard 
deviations become compatible. For this simulation example, rescaling is unlikely
to improve cluster analysis via $k$-means because elongated clusters are not
likely to be eliminated. 
On the other hand, cluster analysis via $\mathcal{M}$-decomposability
involves comparison of pseudo-volumes instead of distances, 
and are therefore {\it invariant} to rescaling of axes.

\subsection{Clustering of Iris Dataset}
Next, we analyze Fisher's Iris dataset via $\mathcal{M}$-decomposability. 
The dataset was obtained from \cite{UCI}.
The dataset consists of 150 four-dimensional data. The four
attribute information given are sepal length, sepal width, petal length and
petal width, all in centimetres. There are altogether three classes,
namely ``Setosa", ``Versicolor" and ``Virginica", in the proportion of
$50:50:50$.

We perform cluster analysis of the dataset via $\mathcal{M}$-decomposability,
 without knowledge of the actual number of classes. At the end of the analysis,
we confirm that there are altogether three classes, 
in the proportion of $50:45:55$. The first $50$ data
coincide with ``Setosa" ($0$ misspecification). For ``Versicolor" and
``Virginica", there are altogether five misspecifications. (Five
``Versicolor" are mislabeled as ``Virginica"). 
The data is depicted graphically in
Fig~$10$ (true class) and Fig~$11$ (estimated class).

Although our analysis results in five cases of misspecifications,
our allocation of ``Versicolor" and ``Virginica" achieves a smaller 
pseudo-volume than the ``true class". Denoting the ``true" classification  
of ``Versicolor" and ``Virginica" by
$\{v_{1}, v_{2}\}$, and our estimation by $\{\hat{v}_{1}, \hat{v}_{2}\}$ 
respectively, our estimation yields
\[
|\Sigma_{\hat{v}_{1}}|^{\frac{1}{2}} + |\Sigma_{\hat{v}_{2}}|^{\frac{1}{2}} 
\approx 0.01563 \,,
\]
as compared to 
\[
|\Sigma_{v_{1}}|^{\frac{1}{2}} + |\Sigma_{v_{2}}|^{\frac{1}{2}} \approx 0.01587
\,.
\]
The pseudo-volume of ``Versicolor" and ``Virginica"
combined into a single class is approximately $0.01799$. 
%Furthermore, refering to Fig~$8.7$ and $8.8$, we see
%that the five misspecified data lie in the vicinity of the ``boundary" between
%``Versicolor" and ``Virginica".

\begin{figure}%[p]
\label{fig:truedata}
\centerline{\includegraphics[width=100mm,clip]{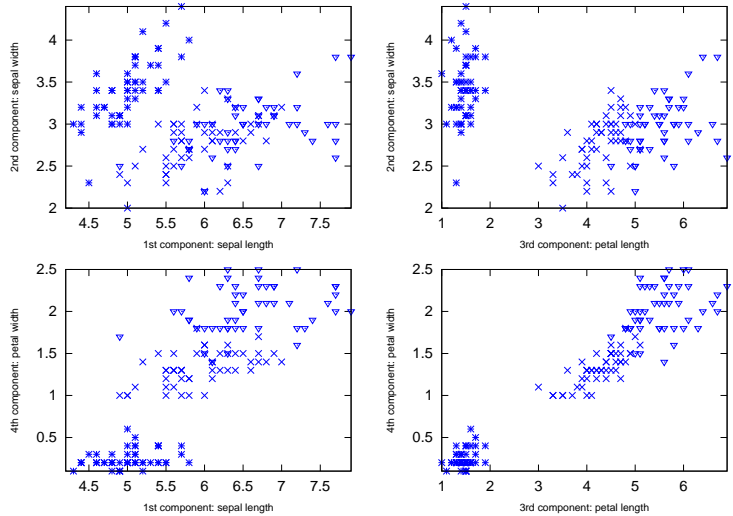}}\vskip2mm
\caption{True Iris data: setosa(asterix), versicolor(cross),
virginica(triangle).} 
%The circles denote data that are misspecified by $\mathcal{M}$-decomposability.}
\end{figure}

\begin{figure}%[p]
\label{fig:estdata}
\centerline{\includegraphics[width=100mm,clip]{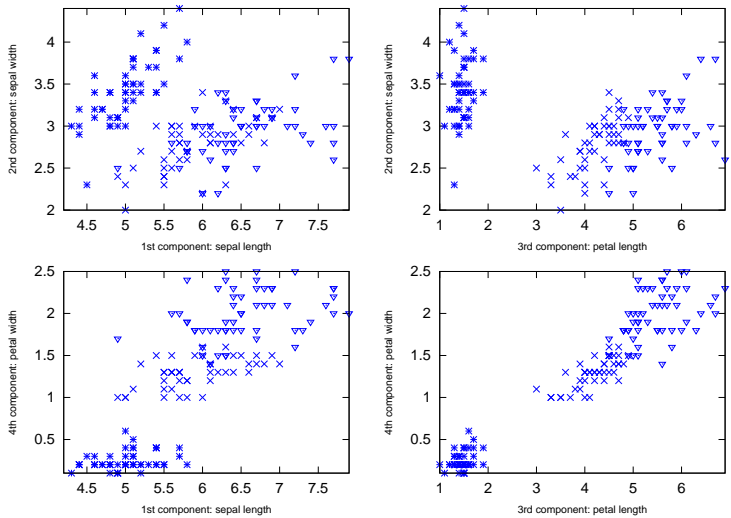}}\vskip2mm
\caption{Iris data recovered via $\mathcal{M}$-decomposability: setosa(asterix),
versicolor(cross), virginica(triangle).}
%The circles denote data that are misspecified by $\mathcal{M}$-decomposability.}
\end{figure}
\newpage

\subsection{Kernel Density Estimation}
\label{sec:densityEstimation}
Density estimation is an important statistical tool that is widely used
in many scientific and engineering fields. Given raw measurements
or data, the task is to recover the unknown density from
which the original data is generated. The problem statement is as
follows. Given $\{X_{1}, \cdots X_{n}\}$, which is generated from 
an unknown distribution with density $f$, the task is to estimate $f$. For
simplicity, we consider only univariate density estimation.

In density estimation, it is usually difficulty to determine quantitatively 
the number of modes in the underlying distribution, just from the given data.
In this respect, Theorem~$\ref{th:KLD}$ can be used for parametric density 
estimation via Gaussian mixtures. 
Besides via Gaussian mixtures, a popular approach to density estimation is via 
the kernel density estimator. The kernel density estimator approach is
non-parametric and is treated in detail in \cite{Scott},
\cite{Silverman}, \cite{WandJones}, \cite{Hardle}. 
The formula for the kernel density estimator, given data 
$\{X_{1}, \cdots X_{n}\}$ is
\begin{equation}
\label{eq:bwf}
\hat{f}(x;b) = (nb)^{-1} \sum_{i=1}^{n} K\{(x-X_{i})/b\} \,,
\end{equation}
see, {\it e.g.} \cite{WandJones}.
Usually $K$ is chosen to be a unimodal density that is symmetric about zero, 
and is called the {\it kernel}. The positive number $b$ is called the {\it
bandwidth}.
Such a formulation ensures that $\hat{f}(x;b)$ is also a density. One property
of the kernel density estimator is that the choice bandwidth is more important
than the choice of the kernel itself. The optimal choice of the bandwidth
ensures that the density estimate becomes optimally smoothed. One popular choice
of the bandwidth is 
\begin{equation}
\label{eq:bw}
b = n^{-\frac{1}{5}} \hat{\sigma} \,,
\end{equation}
where $\hat{\sigma}$ is the sample standard deviation of the given data and $n$
denotes the sample size.
One known problem of the bandwidth given in Eq~(\ref{eq:bw}) is that it works
well for densities that are approximately symmetric unimodal. For multimodal
densities, the bandwidth tends produce an oversmoothed density.

Here, we propose an $\mathcal{M}$-decomposability based algorithm to improve
kernel density estimation. % via the bandwidth given in Eq~(\ref{eq:bw}).
As we are only dealing with the univariate case, we consider just 
the sorted data $F=\{X_{[1]}, \cdots X_{[n]}\}$. Similar to 
Section~$\ref{sec:clustering}$, we perform clustering of $F$ 
via splitting and merging. In one-dimension, the splitting process becomes much
simpler as we just have to find $m$ $(2<m<n-1)$ such that 
$(\sigma_{G} + \sigma_{H})$ is minimized.

For clarity of explanation, we assume that the original data $F$ has two
clusters, and that $G = \{X_{[1]}, \cdots X_{[m]}\}$ and 
$H = \{X_{[m+1]}, \cdots X_{[n]}\}$ are the optimal partition of $F$. We also
have $\sigma_{G} + \sigma_{H} < \sigma_{F}$. As such,
we can expect the density estimation via the weighted mixture of $G$ and $H$
 to be better than that of the original data set. 
Therefore, one may propose an mixture 
kernel density estimator $\hat{f}_{1}$ of $F$ given as follows:
\[
\hat{f}_{1}(x) = \frac{m}{n} \hat{g}(x;b_{g}) + \frac{n-m}{n} \hat{h}(x;b_{h})
\,,
\]
where
\[
b_{g} = m^{-\frac{1}{5}} \hat{\sigma}_{G} \,, \quad
b_{h} = (n-m)^{-\frac{1}{5}} \hat{\sigma}_{H} \,,
\]
and
\[
\hat{g}(x;b_{g}) = (mb_{g})^{-1} \sum_{i=1}^{m} K\{(x-X_{[i]})/b_{g}\} \,,
\]
\[
\hat{h}(x;b_{h}) = \{(n-m)b_{h}\}^{-1} \sum_{i=m+1}^{n} K\{(x-X_{[i]})/b_{h}\} \,.
\]
The original kernel density estimator $\hat{f}$ of $F$ is given in
Eq~(\ref{eq:bwf}).

As an experiment, we generate a sample of size $1000$ from a bimodal density,
with functional form given as
\[
f(x) = \frac{0.2}{\cosh^{2}(x+2.5)} + \frac{0.3}{\cosh^{2}(x-2.5)} \,.
\]
The ``true" density is shown as solid line in Figs~$12$, $13$. 
By simply computing one single bandwith $b$ on
the whole sample set, we obtain a kernel density estimator (computed using
$\hat{f}$). The result is shown as crosses in Fig~$13$. By using
$\mathcal{M}$-decomposability and splitting the data into two clusters, we
obtain a mixture kernel density estimator (computed using $\hat{f}_{1}$). The
result is shown as crosses in Fig~$12$. From Figs~$12$ and $13$, 
it is clear that the kernel density
estimator computed using $\mathcal{M}$-decomposability is closer to
the true density. In this example, we see a pronounced effect of oversmoothing 
(Fig~$13$) for the kernel density estimator with a single 
bandwidth. This is because the original density is bimodal with
modes well separated. The undesirable effect of oversmoothing is alleviated by
implementing $\mathcal{M}$-decomposability.
\begin{figure}%[t]
\centerline{\includegraphics[width=100mm,clip]{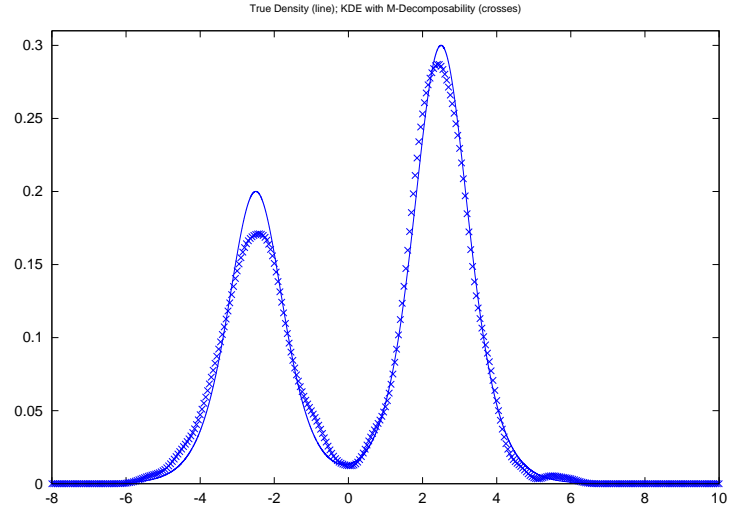}}\vskip2mm
\caption{True density shown as line;
kernel estimate with $\mathcal{M}$-decomposbility shown as crosses.}
\end{figure}

\begin{figure}%[t]
\centerline{\includegraphics[width=100mm,clip]{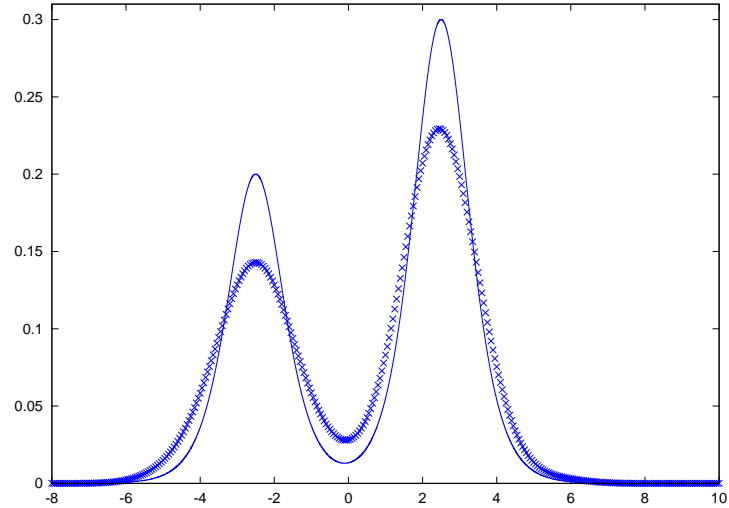}}\vskip2mm
\caption{True density shown as line;
kernel estimate with single bandwidth shown as crosses. 
Comparing with Fig 12, we see that for a multimodal density,
$\mathcal{M}$-decomposability improves kernel density estimation.}
\end{figure}

\section{Conclusion}
\label{sec:conclusion}
In this paper, 
we generalized the notion of $\mathcal{M}$-decom-posability proposed by 
\cite{ChiaNakano1d} to $d$-dimensions, where $d \geq 1$. 
Furthermore, we also broadened the scope of definition of 
$\mathcal{M}$-decom-posability to accomodate any number of mixture components. 
We also derived two theorems pertaining to 
$\mathcal{M}$-decomposability. As a result of 
the first theorem, all elliptical
unimodal densities are $\mathcal{M}$-undecomposable. Consequently, 
any density that is $\mathcal{M}$-decomposable cannot
belong to the class of elliptical unimodal densities, which includes many
general densities, such as Gaussian, Laplace, uniform, logistic, {\it etc.}
The second theorem goes further to say that if a density is
$\mathcal{M}$-decomposable, then it is possible to model the density better 
via a weighted mixture of Gaussian densities. The goodness of fit here is
defined in Kullback-Leibler sense.
$\mathcal{M}$-decomposability is closely related to the modality of probability 
density functions, and hence the theoretical results derived from this paper 
should appeal to theoreticians and practitioners alike. 

We proposed
$\mathcal{M}$-decomposability as a criterion to determine the modality of a
given density, {\it i.e.} if the density is unimodal or multimodal.
%locate modes in a density. 
A practical application is non-parametric cluster analysis. 
Here, one does not need to 
know the parametric model for the underlying clusters. The only
assumption required is that the underlying clusters are approximately elliptical
and unimodal. In this sense, clustering via $\mathcal{M}$-decomposability is 
more flexible and robust than clustering via parametric models or via $k$-means.
Furthermore, we designed a clustering algorithm which automatically determines 
the number of clusters. Our algorithm have been tested on non-Gaussian
cluster examples, as well as the popular Iris dataset.
Another example of application of $\mathcal{M}$-decomposability is density 
estimation. We also devised a scheme to improve kernel density estimation.

Cluster analysis and kernel density estimation are closely related to 
statistical learning. Examples are given in \cite{Hastie}. %further
%scientific and statistical applications of 
Therefore, $\mathcal{M}$-decomposability will also be useful in areas such
as independent component analysis [\cite{Comon}, \cite{Hyvarinen}], machine
learning [\cite{Hand}], {\it etc}. 
Furthermore, as $\mathcal{M}$-decomposability has been 
demonstrated to improve density estimation, it may also be applied to the
improvement of proposal densities in 
%particle filtering [\cite{KotechaDjuric}] 
{\it Markov chain Monte Carlo} (MCMC) methodologies [\cite{RobertCasella}] and 
particle filtering. For example, in \cite{KotechaDjuric}, 
a class of particle filters, called
Gaussian particle filters were introduced. To represent the prior density at each
time-step, the authors generated particles from the Gaussian density fitted to 
the weighted particles representing the previous posterior density. 
Using Theorem~\ref{th:KLD}, the estimation of the prior
density can be improved by fitting a mixture of Gaussian densities to the
weighted particles if necessary, using $\mathcal{M}$-decomposability as 
the criterion to determine the fit. Similarly, in \cite{LeeChia}, 
the authors used Gaussian densities as proposal densities to generate the next 
prior density via MCMC. Using $\mathcal{M}$-decomposability, it is possible to 
improve the proposal densities, which in turn enhances mixing and improves the 
acceptance rates of the sequential MCMC steps.

\section{Appendix}%{Notations and Theorems Used in the Paper}
\subsection{Special Orthogonal Matrices}
%Special orthogonal matrices in $d$-dimensional space ($\mathcal{SO}(d)$) 
%are a class of matrices which satisfy the following
%\[
%A \, A^{T} = {\mathbf I}_{d}, \quad |A| = 1 \,,
%\]
%where ${\mathbf I}_{d}$ denotes the $d$-dimensional identity matrix.
A class of matrices in $d$-dimen-sional space satisfying
\[
A^{-1} = A^{T}, \qquad |A| = 1 %\,,
\]
is given the name {\it special orthogonal matrices}, 
and denoted as $\mathcal{SO}(d)$.
Special orthogonal matrices include all rotation matrices in $d$-dimensional
space. They play an important role in the proof of Lemma~\ref{lm:MinDetCovd}.
The next theorem, which is related to the representation of special orthogonal
matrices, is brought to our attention from
\cite{Bernstein}.
\begin{theorem}%[Farebrother, Wrobel (2002)]
%\begin{thm}%[Farebrother, Wrobel (2002)]
\label{th:SOd}
Let $A \in \mathcal{R}^{d \times d}$, where $d \geq 2$. Then $A \in
\mathcal{SO}(d)$ if and only if there exist $m$ such that $1 \leq m \leq
d(d-1)/2$, $\theta_{1}, \ldots, \theta_{m} \in \mathcal{R}$, and $j_{1}, \ldots,
j_{m}, k_{1}, \ldots, k_{m} \in \{1, \ldots, d\}$ such that
\[
A = \prod_{i=1}^{m} P(\theta_{i},j_{i},k_{i}),
\]
where
\[
P(\theta,j,k) \equiv {\mathbf I}_{d} + (\cos \theta - 1)(E_{j,j} + E_{k,k}) +
(\sin \theta) (E_{j,k} - E_{k,j}) \,.
\]
Here, ${\mathbf I}_{d}$ denotes the $d$-dimensional identity
matrix and 
$E_{i,j}$ denotes the $d \times d$ matrix with one at the $(i,j)$-th element and
zeros everywhere else. 
\end{theorem}
%\end{thm}
The proof is given in \cite{FarebrotherWrobel}.
\begin{remark}
$P(\theta,j,k)$ is a {\it plane} or {\it Givens rotation}.
\end{remark}

\begin{remark}
Theorem~$\ref{th:SOd}$ is an extension of Euler's rotation theorem, which is the
case when $n=3$.
\end{remark}

\subsection{Proof of Lemma~\ref{lm:MinDetCovd}}
\label{sec:proofLmMinDetCovd}
%\begin{proof}
Without loss of generality, we set the mean of $f$ to the origin to simplify 
computations.
Next, note that it is possible to apply a linear transformation to the support
space of $f$, such that the transformed density $f^{w}$ satisfies
\[
\Sigma_{f_{w}} = k_{f} \, {\bf I}_{d} \,, \qquad
|\Sigma_{f_{w}}| = |\Sigma_{f}| \,,
\]
%$M$ to apply to the 
%Next, without loss of generality, we set the covariance matrix of $f$ to be  
%\begin{equation}
%\label{eq:fw}
%\Sigma_{f} = k_{f} \, {\bf I}_{d} \qquad \text{where} \ 
%k_{f} = |\Sigma_{f}|^{\frac{1}{d}} \,,
%\end{equation}
where ${\bf I}_{d}$ denotes the $d$-dimensional identity matrix. 
As a result of the linear transformation, we must also have
\[
\max(f^{w}) = \max(f) \,.
\]
%If Eq~\eqref{eq:fw} does not hold, we simply have to find the linear transformation
%$M$, which applied to $f$, achieves Eq~\eqref{eq:fw}. Such a linear 
%transformation $M$ leaves $|\Sigma_{f}|$, and consequently $\max(f)$, unchanged.
% and we are only interested in the {\it pseudo-volume} of $f$, 
%which is given by the square-root of $|\Sigma_{f}|$.

%This is because it is always possible
%to set the mean of the $f$ to the origin, and then follow by a linear 
%transformation on $f$, without changing either the maximum value of the 
%resultant density, or the determinant of the covariance matrix.

%This is because $\Sigma_{f}$ can be written as 
%\[
%\Sigma_{f} = A^{T} \cdot A \,,
%\]
%and by applying a transformation given by 
%\[
%M = |\Sigma_{f}|^{\frac{1}{2 \, d}} \cdot A^{-1} \,
%\]
%to the density $f$, the covariance matrix of the transformed
%density becomes
%\[
%M^{T} \cdot \Sigma_{f} \cdot M = |\Sigma_{f}|^{\frac{1}{d}} \cdot {\bf I}_{d} \,.
%\] 
%Hence, the transformed density has covariance matrix as a multiple of the
%identity matrix, and pseudo-volume $|\Sigma_{f}|$. Furthermore, since $|M| = 1$,
%the maximum of the transformed density is equal to $\max(f)$.
% para 1

Next, we denote by $u$ the density of the spherical uniform that satisfies
 $\max(u) = \max(f^{w})$. Our goal is then to prove that 
 $|\Sigma_{f^{w}}| \geq |\Sigma_{u}|$, with identity holding if and only if 
 $f^{w}=u$. 
In order to facilitate comparisons of pseudo-volumes of $f^{w}$ and $u$, 
we shall construct a {\it spherical} density $f^{s}$ 
(see Definition~\ref{def:spherical}) that satisfies
\[
\Sigma_{f^{s}} = \Sigma_{f^{w}} \,.
\]
By construction, we have $|\Sigma_{f^{s}}| = |\Sigma_{f^{w}}|$ and
therefore, an equivalent statement of our
goal is $|\Sigma_{f^{s}}| \geq |\Sigma_{u}|$. The steps for the construction of
$f^{s}$ are given in the following paragraph.
%\end{enumerate}

% para 3
%Starting from any $f$ ($\Sigma_{f} \propto {\mathbf I}_{d}$), 
%we construct a spherical density $f^{s}$, which satisfies Item~(\ref{item1}). 
We denote by $f_{j}$ the resultant probability density function when a
rotation operator $R_{j} \in \mathcal{SO}(d)$ is applied onto the support space 
of $f^{w}$. We have
\[
\Sigma_{f_{j}} = \Sigma_{f^{w}} \quad \text{and} \quad
\mu_{f_{j}} = {\bf 0} = \mu_{f} \, .
\]
In other words, the mean and covariance of $f^{w}$ are 
{\it invariant to rotation} if $\Sigma_{f^{w}} \propto {\mathbf I}_{d}$. 
For any rotation operators $R_{i},R_{j} \in \mathcal{SO}(d)$,
any {\em weighted mixture} of $f_{i}$ and $f_{j}$ will again have the same mean 
and covariance matrix. Denoting the mixture by $g$, we have
\[
g({\bf x}) = \alpha f_{i}({\bf x}) + (1-\alpha) f_{j}({\bf x}),
\]
where $0 < \alpha < 1$.  % and $i, j \in \mathcal{N}$. 
The covariance of $g$ is given by
\[
\Sigma_{g} = \alpha \Sigma_{f_{i}} + (1-\alpha) \Sigma_{f_{j}}
+ \alpha (1-\alpha) (\mu_{f_{i}} - \mu_{f_{j}}) (\mu_{f_{i}} -
\mu_{f_{j}})^{T} %\\
	= \Sigma_{f^{w}}.
\]
In two-dimensional space, a rotation operator can be represented as 
\[
R^{\theta} = \left(
		\begin{array}{cc}
			\cos\theta & -\sin\theta \\
			\sin\theta & \cos\theta
		\end{array}
	\right) \, .
\]
From Theorem~\ref{th:SOd}, it is possible to represent any rotation in
$d$-dimensional space as a product of Given's rotations shown below.
\[
R = R_{1}^{\theta_{1}} \cdots R_{D}^{\theta_{D}}
\]
where $D = d \, (d-1) / 2$. 
We are ready to construct $f^{s}$ as follows:
\begin{equation}
\label{eq:fs}
f^{s}({\bf x}) = (\frac{1}{2 \pi})^{D} 
\underbrace{\int_{0}^{2 \pi} \cdots \int_{0}^{2 \pi}}_{\text{$D$ times}}
\, f^{w}(R_{1}^{\theta_{1}} \cdots R_{D}^{\theta_{D}} {\bf x}) \, 
d\theta_{1} \cdots d\theta_{D} \, .
\end{equation}
By construction, $f^{s}$ is the uniform mixture of all possible
rotations of the probability density function $f^{w}$ in $d$-dimensional space.
To show that $\Sigma_{f^{s}} = \Sigma_{f^{w}}$, note that
\[
\begin{split}
\Sigma_{f^{s}} &= \int {\bf x} \, {\bf x}^{T} \, f^{s}({\bf x}) \, d{\bf x} \\
  &= (\frac{1}{2 \pi})^{D} 
\underbrace{\int_{0}^{2 \pi} \cdots \int_{0}^{2 \pi}}_{\text{$D$ times}}
\, 
\underbrace{
\{\int {\bf x} \, {\bf x}^{T} \,
f^{w}(R_{1}^{\theta_{1}} \cdots R_{D}^{\theta_{D}} {\bf x}) \, 
d{\bf x}\} 
}_{A}
\,d\theta_{1} \cdots d\theta_{D} \, .
\end{split}
\] 
The term $A$ is simply the covariance matrix of the transformed density after 
applying rotation operator $R_{1}^{\theta_{1}} \cdots R_{D}^{\theta_{D}}$ to the
support space of $f^{w}$. 
As $\Sigma_{f^{w}}$ is invariant to rotation, we have
%the result remains as $\Sigma_{f}$. Therefore, 
\[
\Sigma_{f^{s}} = (\frac{1}{2 \pi})^{D} \{
\underbrace{\int_{0}^{2 \pi} \cdots \int_{0}^{2 \pi}}_{\text{$D$ times}}
\,d\theta_{1} \cdots d\theta_{D}\} \, \Sigma_{f^{w}} = \Sigma_{f^{w}} \, .
\]
Furthermore, $f^{s}$ must be {\em spherical} %by construction, 
as one can easily verify that $f^{s}(R \, {\bf x}) = 
f^{s}(\bf x)$ for any $R \in \mathcal{SO}(d)$. % \subset \mathcal{O}(d)$.
On top of these, from Eq~$\eqref{eq:fs}$, we have
\begin{equation}
\label{eq:fs<f}
f^{s}({\bf x}) \leq (\frac{1}{2 \pi})^{D} 
\underbrace{\int_{0}^{2 \pi} \cdots \int_{0}^{2 \pi}}_{\text{$D$ times}}
\, \max(f^{w}) \, 
d\theta_{1} \cdots d\theta_{D} = \max(f^{w}) \, .
\end{equation}
We have therefore constructed a spherical density $f^{s}$ whose covariance
matrix is the same as that of $f^{w}$.
Now we are left with proving that $|\Sigma_{f^{s}}| \geq |\Sigma_{u}|$ 
to complete the proof of the lemma.

% para 4
%Next, it is always possible to find $k_{u}>0$ such that 
We express the covariance matrix of $u$ by
$\Sigma_{u} = k_{u} \, {\bf I}_{d}$. Our goal will be accomplished if we
can prove that $k_{f} \geq k_{u}$. From Eq~(\ref{eq:fs<f}), we have
$f^{s}({\bf x}) \leq \max(f) = M_{f}$, and the followings are straightfoward:
\begin{enumerate}
\item $u({\bf x}) \geq f^{s}({\bf x})$ for $| {\bf x} | \leq R$, where 
$u({\bf x}) = M_{f}$ throughout.
\item $u({\bf x}) \leq f^{s}({\bf x})$ for $| {\bf x} | > R$, where 
$u({\bf x}) = 0$ throughout. 
\end{enumerate}
Here, $R$ represents the radius of the spherical uniform $u$. 
Moreover, as $f^{s}$ and $u$ are both spherical and have means centred at the
origin, there exist functions $\tilde{f}^{s}$ and $\tilde{u}$ such that
\[
f^{s}({\bf x}) = \tilde{f}^{s}(|{\bf x}|) = \tilde{f}^{s}(r); \quad 
u({\bf x}) = \tilde{u}(|{\bf x}|) = \tilde{u}(r) \, ,
\]
using Definition~$\ref{def:spherical}$ and representation in the hyperspherical
coordinates. Furthermore, we define 
$h({\bf x}) \equiv f^{s}({\bf x}) - u({\bf x}).$ Note that $h({\bf x})$ is
{\em not} a probability density function as $h({\bf x})$ takes negative values
and 
\begin{equation}
\label{eq:h=0}
\int h({\bf x}) \, d{\bf x} = 0.
\end{equation}
Using the hyperspherical coordinate representation, there must exist a function
$\tilde{h}$ such that $h(|{\bf x}|) = \tilde{h}(r)$, and
\[
%\begin{equation}
%\label{eq:tildeh}
\tilde{h}(r) \ \begin{cases}
                    \leq 0 &\text{for } r \leq R; \\
                    \geq 0 &\text{for } r > R.        
                 \end{cases}
%\end{equation}
\]
Note that $\tilde{h}$ is identically $0$ if and only if $f^{s}=u$, or
equivalently, $f$ is elliptical uniform.  Now, 
\[
\begin{split}
k_{f}-k_{u} &= {\bf e_{1}}^{T}(\Sigma_{f^{s}} - \Sigma_{u}) \, {\bf e_{1}}
\\ 
 &= \int {\bf e_{1}}^{T} {\bf x} {\bf x}^{T} {\bf e_{1}} \{f^{s}({\bf x}) 
 - u({\bf x})\} \, d{\bf x} \\
    &= \int | {\bf e_{1}}^{T} {\bf x} |^{2} \ h({\bf x}) \ d{\bf x} \, .
\end{split}
\]
Here, ${\bf e}_{1}$ is the unit vector parallel to the first axis.
Representation via spherical coordinates yields
\[
\begin{split}
 k_{f}-k_{u} 
	&= \int \cdots \int x_{1}^{2} \, \tilde{h}(r) \, 
	r^{d-1} \, \sin^{d-2}(\phi_{1}) \cdots \sin(\phi_{d-2}) 
	\, dr \, d\phi_{1} \cdots d\phi_{d-1} \\
	&= \int_{0}^{\infty} r^{d+1} \, \tilde{h}(r) \, dr 
	\times \Phi_{1} \times \cdots \times \Phi_{d-1} \, ,
%	\int \cos^{2}(\phi_{1}) \, \sin^{d-2}(\phi_{1}) \, d\phi_{1}
\end{split}
\]
with 
\[
\Phi_{1}=\int_{0}^{\pi} \cos^{2}(\phi_{1}) \, \sin^{d-2}(\phi_{1}) \,
d\phi_{1}, \quad \Phi_{d-1} = 2 \, \pi \, ,
\]
and the rest of $\Phi_{i}$'s ($2 \leq i \leq d-2$) satisfying
\[
\Phi_{i} = \int_{0}^{\pi} \sin^{d-i-1}(\phi_{i}) \, d\phi_{i} \, .
\]
Apparently, all $\Phi_{i}$'s are strictly positive and 
we only need to prove 
\begin{equation}
\label{eq:chebychev}
\int_{0}^{\infty} r^{d+1} \, \tilde{h}(r) \, dr \geq 0  \tag{$*$}
\end{equation}
to arrive at the conclusion that $k_{f} \geq k_{u}$.
Representing Eq~$\eqref{eq:h=0}$ via hyperspherical coordinates, we have
\[
\int_{0}^{\infty} r^{d-1} \, \tilde{h}(r) \, dr 
	\times \int_{0}^{\pi} \sin^{d-2}(\phi_{1}) \, d\phi_{1} 
	\times \Phi_{2} \times \cdots \times \Phi_{d-1} = 0
\]
and therefore
\[
%\begin{equation}
%\label{eq:h=0(s)}
\int_{0}^{\infty} r^{d-1} \, \tilde{h}(r) \, dr = 0 \, .
%\end{equation}
\]
To prove \eqref{eq:chebychev}, we break up the integral into as follows:
\[
\begin{split}
 \int_{0}^{\infty} r^{d+1} \, \tilde{h}(r) \, dr 
 &= \int_{0}^{R} r^{d-1} \, r^{2} \, 
 \underbrace{\tilde{h}(r)}_{\leq 0} \, dr
+ \int_{R}^{\infty} r^{d-1} \, r^{2} \, 
\underbrace{\tilde{h}(r)}_{\geq 0} \, dr \\
 &\geq \int_{0}^{R} r^{d-1} \, R^{2} \, \tilde{h}(r) \, dr
+ \int_{R}^{\infty} r^{d-1} \, R^{2} \, \tilde{h}(r) \, dr \\
 &= R^{2} \times \int_{0}^{\infty} r^{d-1} \, \tilde{h}(r) \, dr = 0 \, .
\end{split}
\]
Equality holds if and only if $\tilde{h}=0$ identically, implying that 
$f^{s}$ is spherical uniform, or in other words, $f$ is elliptical uniform.
This proves $k_{f} \geq k_{u}$ and consequently, we have 
\[
|\Sigma_{f}| = |\Sigma_{f^{s}}| \geq |\Sigma_{u}| \,.
\]

Finally, we need to show that the pseudo-volume of an elliptical uniform density
$u$ with $\max(u) = M_{u}$ is given by
\begin{equation}
\label{eq:drv}
|\Sigma_{u}|^{\frac{1}{2}} = \frac{\Gamma(\frac{d}{2}+1)}
{M_{u} \, \{ \pi \, (d+2) \}^{\frac{d}{2}}} \,.
\end{equation}
We first compute the covariance matrix of an uniform density on the {\it
hypersurface} of the $d$-dimensional sphere. Consider the probability mass
function of a discrete random variable $X$ given below:
\[
f_{X}({\bf x}) = 
\begin{cases}
\frac{1}{2 \, d} & \text{if } {\bf x} =  \pm a \, {\bf e}_{j} \,, \ j = 1,
\ldots, d \,, \\
0 & \text{otherwise.}
\end{cases}
\]
The covariance matrix of the above distribution is computed as
$a^{2} \, {\bf I}_{d} / d$. It is possible to generate an uniform density on the
hypersurface of the $d$-dimensional sphere of radius $a$, by applying rotations
to the discrete random variable given in Eq~\eqref{eq:drv}. Therefore, the
covariance matrix of an uniform density on the hypersurface of a $d$-dimensional
sphere of radius $r$ is $a^{2} \, {\bf I}_{d} / d$. By considering a spherical
uniform density as a continuous mixture of hypersurfaces, we obtain the
covariance matrix of a spherical uniform density with radius $r$ as
\begin{equation}
\label{eq:covOfSphericalUniform}
\Sigma_{u} = \frac{{\bf I}_{d}}{d} \,
\frac{\int_{0}^{r} \, a^{d+1} \, da}
{\int_{0}^{r} \, a^{d-1} \, da}
=\frac{r^{2}}{d+2} \, {\bf I}_{d} \,.
\end{equation}
We therefore obtain the pseudo-volume of a spherical uniform density radius $r$
as
\[
|\Sigma_{u}|^{\frac{1}{2}} = \frac{r^{d}}{(d+2)^{\frac{d}{2}}} \,.
\]
Using the fact that the volume of a $d$-dimensional sphere of
radius $r$ is given by
\[
V = \frac{\pi^{\frac{d}{2}} \, r^{d}}{\Gamma(\frac{d}{2}+1)} \,
\]
we obtain the require pseudo-volume. Hence, the proof 
for Lemma~$\ref{lm:MinDetCovd}$ is complete. \qed
%\end{proof}

\subsection{Proof of Theorem~\ref{th:representationEllipticalUnimodal}}
\label{sec:proofThRepresentation}
%\begin{proof}
We can define the following continuous 
function on non-negative values of $y$, for a given $f$:
\[
q(y) = \int \min\{f({\bf x}), y\} \, d{\bf x} \,.
\]
Then, $q$ is increasing with $q(0) = 0$. If $f$ is unbounded, then $q$ is
strictly increasing for all $y$ with $\lim_{y \rightarrow \infty} q(y) = 1$. If
$f$ is bounded such that $\max(f) = F$, then $q$ is strictly increasing for 
$0 \leq y \leq F$ and $q(y) = 1$ for all $y \leq F$.

We can rewrite $f$ as a sum of two positive functions in the form
\[
f({\bf x}) = f^{(1)}({\bf x}) + f^{(2)}({\bf x}) \,,
\]
where $f^{(1)}({\bf x}) = \min\{f({\bf x}),Y\}$ and $Y$ is positive. For a given
$\epsilon > 0$, it is always possible to choose $Y$ such that
\[
1 - \frac{\epsilon}{4} < \int f^{(1)}({\bf x}) \, d{\bf x} = q(Y) < 1 \,,
\] and therefore 
\[
0 < \int f^{(2)}({\bf x}) \, d{\bf x} < \frac{\epsilon}{4} \,,
\]
because $q$ is continuous ranging between 0 and 1.
The above ``slicing" ensures that the function $f^{(1)}$ is bounded from above
by $Y$. Let $h = Y/n$. Define a set of real numbers $\{r_{n,1}, \ldots, r_{n,n}\}$
by 
\[
r_{n,j} = \sup\{r|p(r^{2}) \geq j \, h\} \,.
\]
Here, $p$ the non-increasing function defined on $\mathcal{R}^{+} \cup \{0\}$
which satisfies $f({\bf x}) = p[({\bf x} - \mu)^{T} \Sigma^{-1} ({\bf x} -
\mu)]$. Setting 
\[
\omega_{n,j} = \frac{r_{n,j}^{d}}{\sum_{i=1}^{n} r_{n,i}^{d}} \,,
\]
we can then construct a density $g_{n}$ such that 
\[
g_{n}({\bf x}) = \sum_{j=1}^{n} \omega_{n,j} \, u_{n,j}({\bf x})\,.
\]
Next rewrite $g_{n}$ as a sum of two positive functions
in the form of
\[
g_{n}({\bf x}) = g^{(1)}_{n}({\bf x}) + g^{(2)}_{n}({\bf x}) \,,
\]
where 
\[
g^{(1)}_{n}({\bf x}) = \sum_{j=1}^{n} r_{n,j}^{d} \, \cdot \, h \, \cdot \, u_{n,j}({\bf x}) \,.
\]
Here, all three functions $g_{n}, g_{n}^{(1)}$ and $g_{n}^{(2)}$ are
proportional to one another. Furthermore, by construction, $g_{n}^{(1)}$ is
dominated everywhere by $f^{(1)}$. We also have
\[
0 \leq f^{(1)}({\bf x}) - g_{n}^{(1)}({\bf x})
 \leq \min\{f({\bf x}),h\} \leq h
\,.
\]
It is therefore possible to choose $n$ (and hence $h$) such that 
\[
\int |g_{n}^{(1)}({\bf x}) - f^{(1)}({\bf x})| \, d{\bf x} 
= \int \{ f^{(1)}({\bf x}) - g_{n}^{(1)}({\bf x}) \} \, d{\bf x}
= q(h) < \frac{\epsilon}{4} \,.
\]

Finally, applying the triangle inequality on integrals, we have
\[
\begin{split}
  &\int |g_{n}({\bf x}) - f({\bf x})| \, d{\bf x} \\
 \leq & \int |g_{n}^{(1)}({\bf x}) - f^{(1)}({\bf x})| \, d{\bf x}
 + \int g_{n}^{(2)}({\bf x}) \, d{\bf x}
 + \int f^{(2)}({\bf x}) \, d{\bf x} \,.
\end{split}
\]
The first and third terms on the right-hand-side of the inequality are both less
than $\epsilon / 4$. The second term is 
\[
\int g_{n}^{(2)}({\bf x}) \, d{\bf x} 
= 1 - \int g_{n}^{(1)}({\bf x}) \, d{\bf x}
< 1 - \int f_{n}^{(1)}({\bf x}) \, d{\bf x} + \frac{\epsilon}{4} =
\frac{\epsilon}{2} \,.
\]
Hence, we arrive at  
\[
\int |g_{n}({\bf x}) - f({\bf x})| \, d{\bf x} 
< \epsilon \,,
\]
completing the proof of Theorem~\ref{th:representationEllipticalUnimodal}. \qed
%\end{proof}

\section*{Acknowledgements}
This paper results from the Ph.D. work of Nicholas Chia at the Institute of
Statistical Mathematics. Nicholas Chia would like to express his gratitude to
John Copas, Shinto Eguchi, Katuomi Hirano, Satoshi Kuriki, Kunio Shimizu and 
Yoshiyasu Tamura for their kind and helpful advices. Bill Farebrother 
graciously supplied the proof of Theorem~\ref{th:SOd}, which was crucial to the 
proof of Lemma~\ref{lm:MinDetCovd} and subsequently Theorem~\ref{th:maind}.
Nicholas Chia dedicates this paper to the memory of Taichi Morichika, who is
outlived only by his creativity and passion in research.

%

%% Format for Journal Reference
%{Author} (year). {article's title},
%\textit{Journal}, \textbf{Volume}, {first
%page number}--{last page number}.

%% Format for Book
%{Author} (year). {article's title},
%\textit{Book's title} (eds.\ {Editor}),
%{first page number}--{last page number},
%{publisher}, {published place}.

%Ibragimov, I. A., 
%On the composition of unimodal distributions, 
%Theor. Probability Appl., 1 (1956), 255-260.

%Anderson, T. W. (1955). 
%The integral of a symmetric unimodal function over a symmetric convex
%set and some probability inequalities. 
%Proc.Amer .Math.Soc . 6 170-176.

\end{document}